\documentclass[aps,prl,reprint,twocolumn,color,superscriptaddress,floatfix]{revtex4-2}
\usepackage[english]{babel} 
\usepackage{amssymb}
\usepackage{amsmath}
\usepackage{txfonts}
\usepackage{mathdots}
\usepackage[dvips]{graphicx}
\usepackage{epsfig}
\usepackage{graphicx}
\usepackage{array}
\usepackage{amssymb}
\usepackage{amsfonts}
\usepackage{amsmath}
\usepackage{mathrsfs}
\usepackage{booktabs}
\usepackage{threeparttable}
\usepackage{multirow}
\usepackage{subfigure}
\usepackage{epsfig}
\usepackage{threeparttable}
\usepackage{chngpage}
\usepackage{float}
\usepackage{xcolor}
\usepackage{bm}
\usepackage{graphicx}
\usepackage[toc]{appendix}
\usepackage{tabularx}
\usepackage{graphicx}
\usepackage{adjustbox}
\usepackage{comment}
\usepackage{hyperref}
\hypersetup{
    unicode=false,          
    pdftoolbar=false,        
    pdfmenubar=true,        
    pdffitwindow=false,     
    pdfstartview={FitH},    
    pdftitle={},    
    pdfauthor={Authors},     
    pdfsubject={},   
    pdfcreator={},   
    pdfproducer={}, 
    pdfkeywords={quantum many-body scars} {superconducting processor} {quantum state tomography}, 
    pdfnewwindow=true,      
    colorlinks=true,       
    linkcolor=black,          
    citecolor=blue,        
    filecolor=magenta,      
    urlcolor=blue           
}

\newcommand{\beginsupplement}{%
    \setcounter{table}{0}
    \renewcommand{\thetable}{S\arabic{table}}%
    \setcounter{figure}{0}
    \renewcommand{\thefigure}{S\arabic{figure}}%
    \setcounter{equation}{0}
    \renewcommand{\theequation}{S\arabic{equation}}%
   }

\begin{document}

\title{Many-body Hilbert space scarring on a superconducting processor}

\author{Pengfei Zhang} \thanks{These three authors contributed equally}
\affiliation{Department of Physics, ZJU-Hangzhou Global Scientific and Technological Innovation Center, Interdisciplinary Center for Quantum Information, and Zhejiang Province Key Laboratory of Quantum Technology and Device, Zhejiang University, Hangzhou 310027, China}

\author{Hang Dong} \thanks{These three authors contributed equally}
\affiliation{Department of Physics, ZJU-Hangzhou Global Scientific and Technological Innovation Center, Interdisciplinary Center for Quantum Information, and Zhejiang Province Key Laboratory of Quantum Technology and Device, Zhejiang University, Hangzhou 310027, China}

\author{Yu Gao} \thanks{These three authors contributed equally}
\affiliation{Department of Physics, ZJU-Hangzhou Global Scientific and Technological Innovation Center, Interdisciplinary Center for Quantum Information, and Zhejiang Province Key Laboratory of Quantum Technology and Device, Zhejiang University, Hangzhou 310027, China}

\author{Liangtian Zhao}
\affiliation{Institute of Automation, Chinese Academy of Sciences, Beijing 100190, China}

\author{Jie Hao}
\affiliation{Institute of Automation, Chinese Academy of Sciences, Beijing 100190, China}

\author{Jean-Yves Desaules}
\affiliation{School of Physics and Astronomy, University of Leeds, Leeds LS2 9JT, UK}

\author{Qiujiang Guo}
\affiliation{Department of Physics, ZJU-Hangzhou Global Scientific and Technological Innovation Center, Interdisciplinary Center for Quantum Information, and Zhejiang Province Key Laboratory of Quantum Technology and Device, Zhejiang University, Hangzhou 310027, China}
\affiliation{Alibaba-Zhejiang University Joint Research Institute of Frontier Technologies, Hangzhou 310027, China}

\author{Jiachen Chen}
\affiliation{Department of Physics, ZJU-Hangzhou Global Scientific and Technological Innovation Center, Interdisciplinary Center for Quantum Information, and Zhejiang Province Key Laboratory of Quantum Technology and Device, Zhejiang University, Hangzhou 310027, China}

\author{Jinfeng Deng}
\affiliation{Department of Physics, ZJU-Hangzhou Global Scientific and Technological Innovation Center, Interdisciplinary Center for Quantum Information, and Zhejiang Province Key Laboratory of Quantum Technology and Device, Zhejiang University, Hangzhou 310027, China}

\author{Bobo Liu}
\affiliation{Department of Physics, ZJU-Hangzhou Global Scientific and Technological Innovation Center, Interdisciplinary Center for Quantum Information, and Zhejiang Province Key Laboratory of Quantum Technology and Device, Zhejiang University, Hangzhou 310027, China}

\author{Wenhui Ren}
\affiliation{Department of Physics, ZJU-Hangzhou Global Scientific and Technological Innovation Center, Interdisciplinary Center for Quantum Information, and Zhejiang Province Key Laboratory of Quantum Technology and Device, Zhejiang University, Hangzhou 310027, China}

\author{Yunyan Yao}
\affiliation{Department of Physics, ZJU-Hangzhou Global Scientific and Technological Innovation Center, Interdisciplinary Center for Quantum Information, and Zhejiang Province Key Laboratory of Quantum Technology and Device, Zhejiang University, Hangzhou 310027, China}

\author{Xu Zhang}
\affiliation{Department of Physics, ZJU-Hangzhou Global Scientific and Technological Innovation Center, Interdisciplinary Center for Quantum Information, and Zhejiang Province Key Laboratory of Quantum Technology and Device, Zhejiang University, Hangzhou 310027, China}

\author{Shibo Xu}
\affiliation{Department of Physics, ZJU-Hangzhou Global Scientific and Technological Innovation Center, Interdisciplinary Center for Quantum Information, and Zhejiang Province Key Laboratory of Quantum Technology and Device, Zhejiang University, Hangzhou 310027, China}

\author{Ke Wang}
\affiliation{Department of Physics, ZJU-Hangzhou Global Scientific and Technological Innovation Center, Interdisciplinary Center for Quantum Information, and Zhejiang Province Key Laboratory of Quantum Technology and Device, Zhejiang University, Hangzhou 310027, China}

\author{Feitong Jin}
\affiliation{Department of Physics, ZJU-Hangzhou Global Scientific and Technological Innovation Center, Interdisciplinary Center for Quantum Information, and Zhejiang Province Key Laboratory of Quantum Technology and Device, Zhejiang University, Hangzhou 310027, China}

\author{Xuhao Zhu}
\affiliation{Department of Physics, ZJU-Hangzhou Global Scientific and Technological Innovation Center, Interdisciplinary Center for Quantum Information, and Zhejiang Province Key Laboratory of Quantum Technology and Device, Zhejiang University, Hangzhou 310027, China}

\author{Bing Zhang}
\affiliation{Alibaba-Zhejiang University Joint Research Institute of Frontier Technologies, Hangzhou 310027, China}

\author{Hekang Li}
\affiliation{Department of Physics, ZJU-Hangzhou Global Scientific and Technological Innovation Center, Interdisciplinary Center for Quantum Information, and Zhejiang Province Key Laboratory of Quantum Technology and Device, Zhejiang University, Hangzhou 310027, China}
\affiliation{Alibaba-Zhejiang University Joint Research Institute of Frontier Technologies, Hangzhou 310027, China}

\author{Chao Song}
\affiliation{Department of Physics, ZJU-Hangzhou Global Scientific and Technological Innovation Center, Interdisciplinary Center for Quantum Information, and Zhejiang Province Key Laboratory of Quantum Technology and Device, Zhejiang University, Hangzhou 310027, China}
\affiliation{Alibaba-Zhejiang University Joint Research Institute of Frontier Technologies, Hangzhou 310027, China}

\author{Zhen Wang}
\affiliation{Department of Physics, ZJU-Hangzhou Global Scientific and Technological Innovation Center, Interdisciplinary Center for Quantum Information, and Zhejiang Province Key Laboratory of Quantum Technology and Device, Zhejiang University, Hangzhou 310027, China}
\affiliation{Alibaba-Zhejiang University Joint Research Institute of Frontier Technologies, Hangzhou 310027, China}

\author{Fangli Liu}
\affiliation{QuEra Computing Inc., Boston, Massachusetts 02135, USA}

\author{Zlatko Papi\'{c}}
\affiliation{School of Physics and Astronomy, University of Leeds, Leeds LS2 9JT, UK}

\author{Lei Ying}\email{leiying@zju.edu.cn}
\affiliation{Department of Physics, ZJU-Hangzhou Global Scientific and Technological Innovation Center, Interdisciplinary Center for Quantum Information, and Zhejiang Province Key Laboratory of Quantum Technology and Device, Zhejiang University, Hangzhou 310027, China}
\affiliation{Alibaba-Zhejiang University Joint Research Institute of Frontier Technologies, Hangzhou 310027, China}

\author{H. Wang}\email{hhwang@zju.edu.cn}
\affiliation{Department of Physics, ZJU-Hangzhou Global Scientific and Technological Innovation Center, Interdisciplinary Center for Quantum Information, and Zhejiang Province Key Laboratory of Quantum Technology and Device, Zhejiang University, Hangzhou 310027, China}
\affiliation{Alibaba-Zhejiang University Joint Research Institute of Frontier Technologies, Hangzhou 310027, China}

\author{Ying-Cheng Lai}\email{Ying-Cheng.Lai@asu.edu}
\affiliation{School of Electrical, Computer and Energy Engineering, and Department of Physics, Arizona State University, Tempe, Arizona 85287, USA}

\date{\today}

\begin{abstract}
Quantum many-body scarring (QMBS) -- a recently discovered form of weak ergodicity breaking in strongly-interacting quantum systems -- presents opportunities for mitigating thermalization-induced decoherence in quantum information processsing. However, the existing experimental realizations of QMBS are based on kinetically-constrained systems where an emergent dynamical symmetry ``shields" such states from the thermalizing bulk of the spectrum. Here, we experimentally realize a distinct kind of QMBS phenomena by  approximately decoupling a part of the many-body Hilbert space in the computational basis. Utilizing a programmable superconducting processor with 30 qubits and tunable couplings, we realize Hilbert space scarring in a non-constrained model in different geometries, including a linear chain as well as a quasi-one-dimensional comb geometry. By performing full quantum state tomography on 4-qubit subsystems, we provide strong evidence for QMBS states by measuring qubit population dynamics, quantum fidelity and entanglement entropy following a quench from initial product states.
Our experimental findings broaden the realm of QMBS mechanisms and pave the way to exploiting correlations in QMBS states for applications in quantum information technology.
\end{abstract}
\maketitle

\begin{figure*}
\centering
\includegraphics[width=\linewidth]{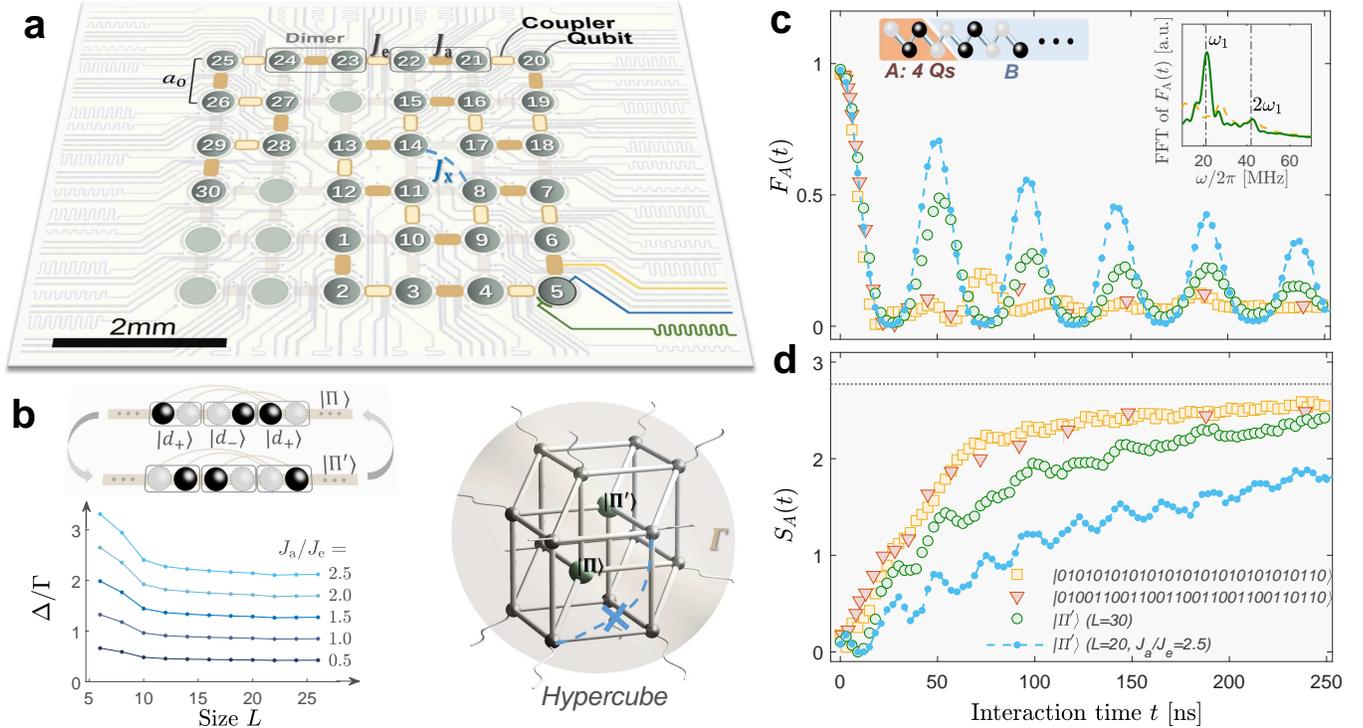}
\caption{  {\bf Experimental setup and identification of QMBS states via quantum state tomography.}
(a) Experimental superconducting circuit with qubits and couplers in a square geometry. Light gray dashed rectangles represent dimers that constitute the chain with intra coupling $J_\mathrm{a}$, inter coupling $J_\mathrm{e}$, and small cross coupling $J_\mathrm{x}$.
(b) Left upper: A schematic illustration of the dynamics of the collective dimer states $|\Pi\rangle$ and $|\Pi^\prime\rangle$.
Left lower: The ratio $\Delta/\Gamma$ as a function of the system size $L$ for different ratios of $J_\mathrm{a}/J_\mathrm{e}$ and for a fixed value of $J_\mathrm{x}$ listed in Fig.~\ref{sfig:Jx} of the Supplementary Material.
Right: a $4$-dimensional hypercube in the Hilbert space.
(c)-(d): Quantum state tomography for the four-qubit fidelity and
entanglement entropy in a $30$-qubit chain for initial
thermalizing states (red and orange) and a QMBS state (green). Blue dot-dashed
curves stand for the dynamics of state $|\Pi^\prime\rangle$ with $L=20$ and a higher
coupling ratio $J_\mathrm{a}/J_\mathrm{e}=2.5$.
(c) Fidelity of the subsystem $A$ as a function of interaction time. The upper
left inset illustrates the partition of the many-body system into subsystems
${A}$ and ${B}$, where the former contains $4$ qubits for measuring the
entanglement entropy. The right inset shows the Fourier transformation
amplitude of the four-qubit fidelity and $\omega_1/2\pi\approx 21$ MHz.
(d) Time evolution of the entanglement entropy ${S}_{A}$.
The dotted grey line represents the thermal value $4 \ln 2$.
}
\label{fig:schematic}
\end{figure*}

\section*{Introduction}

Strongly-coupled quantum systems provide a wealth of opportunities for fundamental physics as well as practical applications that utilise quantum entanglement, which is naturally abundant in them~\cite{LJLN:2010,GAN:2014,MRQM:2021,AABR:2019}. However, the majority of such systems, even if they are perfectly isolated from the external world, undergo chaotic dynamics which gives rise to quantum information scrambling~\cite{SBSH:2016,XCZZ:2018,LFSL:2019,MRQM:2021,MLBC:2021}. The scrambling invariably results in quantum thermalization -- the process described by the so-called Eigenstate Thermalization Hypothesis
(ETH)~\cite{Deutsch:1991,Srednicki:1994,RDO:2008,KTLR:2016}. Thermalization represents a major challenge that needs to be overcome when such systems are used for practical applications. Thus, developing methods to defy the ETH so as to achieve long-lived-dynamic states, thereby preserving quantum information, has become an important goal of quantum sciences~\cite{BOLK:2021}.

Recent discovery of mechanisms for weakly breaking the ETH, such as quantum many-body scarring (QMBS)~\cite{SAP:2021,MoudgalyaReview}, have opened the door to delaying thermalization by preparing the system in special initial states. This route offers more flexibility for designing non-ergodic dynamics than, for example, fine tuning the couplings of the systems to make it integrable, and it avoids the need to strongly disorder the system to drive it into a many-body localized phase~\cite{NH:2015,AABS:2019,GCLX:2021,GCSS:2021}. Because of their ability to suppress thermalization, QMBS states are expected to be useful for storing quantum information~\cite{SAP:2021}, generating Greenberger–Horne–Zeilinger (GHZ) entangled state~\cite{OLKS:2019} and in quantum-enhanced sensing~\cite{Dooley:2021}.
However, while there has been a proliferation of theoretical studies of QMBS in a variety of models~\cite{SM:2017, MRB:2018, OCMC:2019, SI:2019, SYK:2020, MM:2018, MB:2020, MHSR:2020, MRB:2020, VMS:2020, LCH:2020,  HDGC:2020, MHSR:2020, LMPC:2020, Zhao2020, KMH:2020, MNSS:2020, ODea2020, SVLSDG:2021, WSSN:2021, DHTP:2021, RLF:2021},
the experimental realizations of QMBS remain in short supply. The existing QMBS experiments remain focused on a single model with kinetic constraints -- the so-called PXP
model~\cite{FendleySachdev,Lesanovsky2012}, which has been effectively realized using Rydberg atoms~\cite{BSKL:2017,BOLK:2021}
and ultracold bosons in optical lattices~\cite{Su2022}. More recently, ultracold lithium-7 atoms in an optical lattice, which realize the Heisenberg spin model, have been explored as a host of non-thermalizing helix states, reminiscent of QMBS~\cite{Jepsen2021}.


In this article, we report the experimental observation of a new class of QMBS states on a superconducting (SC) processor.
In contrast to previous realizations in kinetically constrained Rydberg atom arrays, we design QMBS by weakly decoupling one part of the Hilbert space in the computational basis. Our approach is inspired by the topological structure of the Su-Schrieffer-Heeger model of polyacetylene~\cite{SSHModel}, which we utilise to create a nearly decoupled subspace with the structure of the hypercube graph. This subspace gives rise to emergent QMBS phenomena, including many-body revivals from special initial states residing in the hypercube, as well as the band of scarred eigenstates. At the same time, the entire system thermalizes due to weaker cross couplings between neighboring qubits. One of the advantages of our SC platform is the tunable $XY$ coupling between qubits (see Methods section for more details)
on a 6 by 6 square lattice configuration, which enables us to emulate many-body systems with both one-dimensional (1D) and quasi-1D systems with comb shape. We investigate circuits of up to $30$ qubits and $29$ couplers, with the Hilbert space dimension  $C(30,15)=155,117,520$ -- far beyond the limits of classical simulation.
Measurements of population dynamics and quantum state tomography for entanglement entropy and quantum fidelity provide strong evidence of the emergence of robust QMBS states, as we demonstrate by directly comparing their slow dynamics against conventional thermalizing states.  Our realization of a new QMBS paradigm in a solid state SC platform paves the way to a systematic exploration of scarring and other forms of ergodicity breaking in systems with highly tunable interactions extending beyond one spatial dimension.

\section*{Experimental setup and observation of many-body scar states via quantum state tomography}

Our experiment utilizes a two-dimensional SC qubit array~\cite{AABR:2019,WBCC:2021}, shown in Fig.~\ref{fig:schematic}(a), which features high density
integration and high degree of controllability over local couplings~\cite{KKYO:2019,BGGW:2021}, allowing to emulate different models in a single device.
We first consider the ``snake"-like qubit layout in Fig.~\ref{fig:schematic}(a). This layout exploits the structure of the Su-Schrieffer-Heeger
chain~\cite{SSHModel}, where the intra-dimer coupling
$J_{i,i+1}/2\pi=J_\mathrm{a}/2\pi \simeq -9$ MHz with $i\in \mathrm{odd}$
is slightly stronger than the inter-dimer coupling
$J_{i,i+1}/2\pi=J_\mathrm{e} /2\pi \simeq -6$ MHz with $i\in \mathrm{even}$.

In the limit $J_\mathrm{a} \gg J_\mathrm{e}$, each dimer behaves as a nearly free two-level system, hence the Hilbert space of the SC qubit system has the structure of the hypercube graph -- see Fig.~\ref{fig:schematic}(b). Such a system supports quantum revivals but they are essentially of a single-particle origin. When $J_\mathrm{a}$ and $J_\mathrm{e}$ are comparable in magnitude, we are in the regime of the SSH model where quench dynamics from fully polarized and N\'eel initial states has recently been investigated in Refs.~\cite{Jafari2017, Najafi2019}. While the N\'eel state does not display persistent revivals, we will show below that it is possible to identify, based on the hypercube structure, other initial states that \emph{do} exhibit quantum revivals in moderate systems with $10$-$20$ qubits. In order to show that these are \emph{bona fide} QMBS states, we will confirm they are robust to both the increase in system size as well as the breaking of integrability that allows the system to thermalize from most other initial states. Thermalization is our experimental setup is naturally driven by the cross couplings $J_\mathrm{x}/2\pi \in [0.3,1.2]$ MHz -- the
couplings between two next nearest neighbor qubits with a physical separation
distance  $a_{ij}=\sqrt{2}a_0$, where $a_{0}\approx0.8$ mm is the
separation distance of two nearest neighbor qubits. These couplings  break the reflection symmetries of the circuit and make the system thermalize. For example, exact diagonalization confirms that the energy level spacings follow the Wigner-Dyson distribution with the level-statistics parameter $\langle r \rangle\approx 0.53$ for large symmetry-resolved sectors -- see Supplementary Material (SM).  

Now we identify QMBS states based on their special location in the hypercube. Recall that each dimer has four states:
$|\mathrm{d}_0\rangle=|00\rangle$, $|\mathrm{d}_1\rangle=|11\rangle$,
$| \mathrm{d}_+\rangle= |10\rangle$, and $|\mathrm{d}_-\rangle=| 01\rangle$.
At half filling, i.e., with the number of photons $N$ equal to half the total number of qubits $L$,  a special class of dimerized states in the computational basis states can be identified. These states all have a single photon in each dimer (i.e they only have $| \mathrm{d}_+\rangle$ or $| \mathrm{d}_-\rangle$), and the connectivity between them in the Hamiltonian forms $N=L/2$-dimensional hypercube with equal weight edges.
Among these, a pair of collective states $|\Pi\rangle=|\mathrm{d}_+\mathrm{d}_-\mathrm{d}_+\mathrm{d}_-\cdots \rangle$
and $|\Pi^\prime\rangle=|\mathrm{d}_-\mathrm{d}_+\mathrm{d}_-\mathrm{d}_+\cdots \rangle$, illustrated in Fig.~\ref{fig:schematic}(b), play a special role. These are located on opposite corners of the hypercube and have the unique property of only having intra-dimer couplings (with the exception of the small $J_\mathrm{x}$ couplings).  This helps prevent the information in states $|\Pi \rangle$ and $|\Pi^\prime\rangle$ from rapidly
leaking into the thermalizing part of the Hilbert space, with the other product states in the hypercube playing the role of a ``buffer'' area. The robustness of $|\Pi \rangle$ and
$|\Pi^\prime\rangle$ is due to the collective many-body effect and enhanced
by the structure-induced potential ($J_\mathrm{a}>J_\mathrm{e}$).
The hypercubic structure is robust and naturally it does not contain any cross
coupling. Indeed, while the other parts of the Hilbert space are frustrated by the irregular $J_\mathrm{x}$ couplings, no two states within the hypercube are linked by them. They contribute to the leakage of the wavefunction out of the hypercube but do not affect the dynamics within it.

\begin{figure*}
\centering
\epsfig{figure=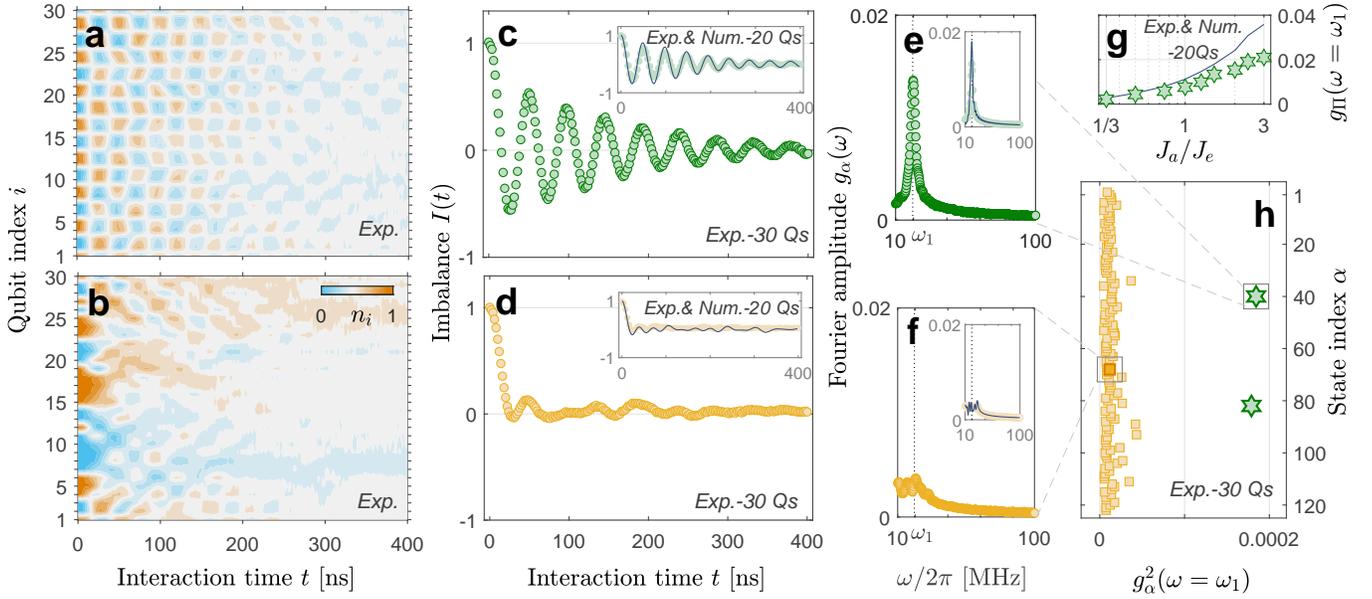,width=\linewidth}
\caption{ {\bf Experimentally observed qubit dynamics.}
(a)-(b): Contour diagrams of the experimental qubit population as a function of
the interaction time for a QMBS and a rapid thermalizing state, respectively.
(c)-(d): Generalized imbalance $I(t)$ extracted from plots (a)-(b) as a function of the interaction time. Insets:
imbalance dynamics from experiments (dots or circles) and numerical
simulations (solid curves) in a $20$-qubit chain.
(e)-(f): The Fourier transformation amplitude of the imbalance in
(c)-(d), which characterizes the squared overlap between the initial states and the energy eigenstates. The time window for the fast Fourier transform is extended to $4\ \mu$s with zero padding.
(g) Fourier peak as a function of the coupling ratio
$J_\mathrm{a}/J_\mathrm{e}$ in a chain of $L=20$ from experimental
measurements (green hexagrams) and numerical simulations (dashed curve).
(h) The squared Fourier amplitude $g_\alpha^2(\omega=\omega_1 )$ for  $|\alpha\rangle$ for $120$ randomly chosen initial product states,
including two QMBS states (green hexagrams) that are clearly stand out from the rest of thermalizing product states (yellow squares). The simulation parameter values
are $J_\mathrm{a}/2\pi=-9.3$ MHz, $J_\mathrm{e}/2\pi=-6.1$ MHz and
$J_\mathrm{x}/2\pi \in [0.3,1.2]$ MHz.}
\label{fig:evolution}
\end{figure*}

The sum of the hypercubic thermal couplings (inter dimer and cross couplings) gives the decay rate $\Gamma$ of the hypercube to the thermalized parts. The
summation of intra hypercubic couplings $\Delta$ is given by the number
of hypercubic edges, i.e., $\Delta=N 2^{N-1} J_\mathrm{a}$. The ratio between
the sums of the inter and intra hypercubic couplings $\Delta/\Gamma$ converges
to a finite value for different values of $J_\mathrm{a}/J_\mathrm{e}$ (see in Fig.~\ref{fig:evolution} (b)), which shows that the hypercube is not trivially disconnected from the rest of the Hilbert space.


With the high-precision control and readouts of our SC processor, we are able to perform tomography measurements to obtain the non-diagonal
elements of the reduced density matrix $\rho_A$, which determine the time evolution of the fidelity of the subsystem $F_A(t)$ and the entanglement entropy $S_A(t)$. The complexity of such measurements grows rapidly with the size of the subsystem $A$ and below we consider $A$ to be four qubits, as schematically illustrated in the inset of Fig.~\ref{fig:schematic}(c). The data points in Fig.~\ref{fig:schematic}(c) give,
for a $30$-qubit chain, the time evolution of the first four-quit fidelity for
a collective state $|\Pi^\prime\rangle$ and that of a typical thermalizing state. The
fidelity of the QMBS state exhibits remarkable revivals with the period of
about $50$ ns and the peak value of the first revival can be as high as $0.5$,
while no such revivals occur for the thermalizing state.

We emphasize that although we consider a relatively small subsystem of fixed size here, numerical simulations confirm that the behavior of the subsystem fidelity closely mirrors that of the fidelity for the pure state $|\Psi(t)\rangle$ of the entire system (see SM). The revivals are enhanced by increasing the ratio $J_\mathrm{a}/J_\mathrm{e}$, which controls the coupling of the hypercube to the rest of the Hilbert space.  As we emphasized above, in the extreme limit $J_\mathrm{a} \gg J_\mathrm{e}$, the dimers are only weakly interacting with each other and the revival dynamics has a single-particle origin. Furthermore, the revivals of the states $|\Pi\rangle$ and $|\Pi^\prime\rangle$ become weak if $L/2$ is an odd integer with a periodic boundary condition, which is experimentally confirmed (see SM).


In Fig.~\ref{fig:schematic}(d) we measure the time evolution of the entanglement entropy for both QMBS and conventional thermalizing states. The entanglement entropy is defined as the von Neumann entropy $S_\mathrm{A} = - \mathrm{Tr} [\rho_{A} \mathrm{log}\rho_{A}]$, where $\rho_{A}$ is the reduced density matrix of the subsystem $A$.
Fig.~\ref{fig:schematic}(d) shows the time evolution of $S_A$ for the state
$|\Pi\rangle$ as well as a random thermalizing state. The scarred dynamics leads to a slow growth of entanglement entropy, superposed with oscillations whose frequency is twice that of fidelity revivals in Fig.~\ref{fig:schematic}(c). By contrast, for the thermalizing initial state, the entropy rises rapidly towards the value $4\mathrm{ln}(2)$, which is approximately the maximum Page entropy for this subsystem.
In addition, the scar features of the states $|\Pi\rangle$ and $|\Pi^\prime\rangle$ can be enhanced
by increasing the coupling ratio $J_\mathrm{a}/J_\mathrm{e}$, as illustrated by the suppressed entropy growth in Fig.~\ref{fig:schematic}(d).

\section*{Qubit dynamics beyond the limit of classical simulations}

To further probe the dynamics at the level of individual qubits, we measure the generalized population imbalance defined as ${I}(t)=(1/L)\sum_i^L\langle\mathcal{S}_i^z(0)\rangle\langle\mathcal{S}_i^z(t)\rangle$. The imbalance is determined by the overlaps $|\langle E_n | \alpha\rangle|^2$ of energy eigenstates $|E_n\rangle$ with the initial state $|\alpha\rangle$ and the phase factors $\exp(-i(E_n-E_m) t /\hbar)$, where $m,n$ are eigenstate indices. For a thermalizing initial state, the phases are essentially random and the initial state has roughly equal support on all energy eigenstates. Thus, any imbalance present in the initial state will rapidly diffuse to a value exponentially small in the system size and it cannot be detected via local operators at late times. By contrast, a QMBS initial state has appreciable overlap only on a \emph{few}  eigenstates with phases set to integer multiples of a single dominant frequency. This allows a QMBS state to display a persistent quantum revival even at relatively late times.

\begin{figure}[htb]
\centering
\epsfig{figure=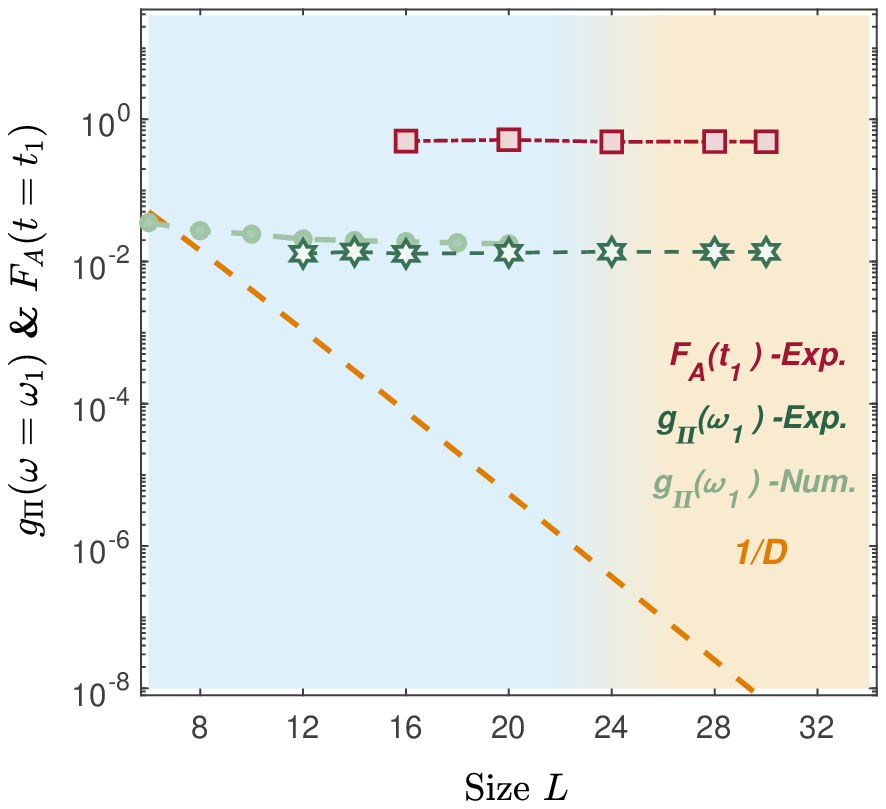,width=3in}
\caption{{\bf Scaling behavior} of the
first revival peak $F_A(t_1\approx52\mathrm{ns})$ of the subsystem $A$,
Fourier amplitude $g_\Pi(\omega_1)$, and the inverse Hilbert space dimension
$1/D$ versus the system size for $J_\mathrm{a}/2\pi\approx-9$ MHz and
$J_\mathrm{e}/2\pi\approx-6$ MHz. The light blue area denotes the regime
where classical simulations using the exact diagonalization method are feasible.}
\label{fig:scaling}
\end{figure}

\begin{figure*}
\centering
\epsfig{figure=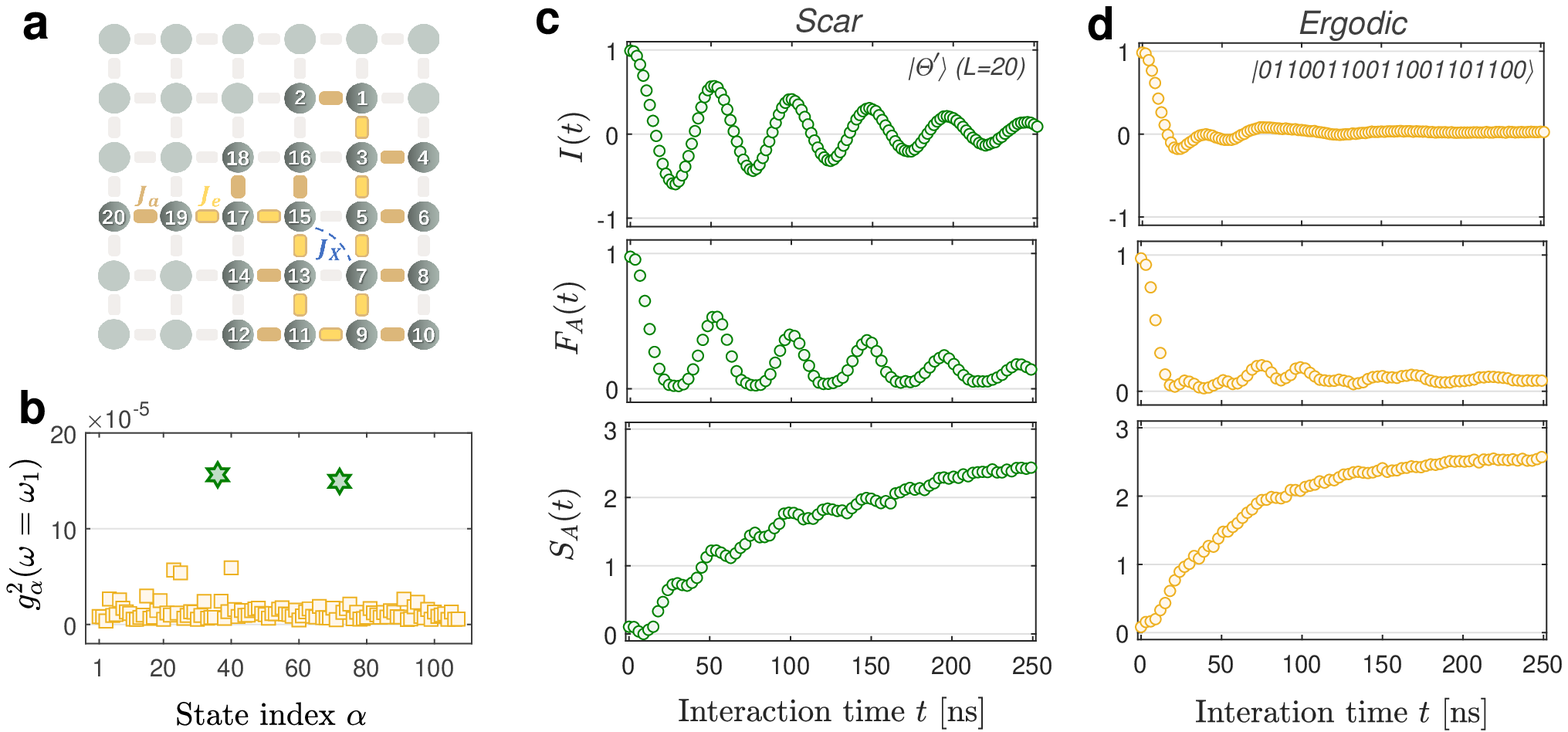,width=\linewidth}
\caption{{\bf QMBS states in a comb tensor system.}
(a) The comb tensor topology in the superconducting processor with $L=20$
qubits. (b) Squared Fourier amplitude $g_\alpha^2(\omega=\omega_1)$ for
randomly chosen initial product states. The QMBS states (green hexagrams) are
characteristically distinct from the conventional thermalizing product states.
(c,d) Dynamics of the imbalance (upper), four-qubit fidelity (middle), and
four-qubit entanglement entropy (lower) for the state $|\Theta^\prime\rangle$
(c) and a randomly choose product state (d). The couplings are tuned to
$J_\mathrm{a}/2\pi\approx-9$ and $J_\mathrm{e}/2\pi\approx-6$ MHz. All data
from experimental measurements. }
\label{fig:comb}
\end{figure*}

The evolution of population imbalance for individual qubits in a 30-qubit chain is shown in Figs.~\ref{fig:evolution}(a) and \ref{fig:evolution}(b), which contrast the evolution of a QMBS state and a typical thermalizing state. QMBS exhibits remarkable oscillations which are absent in the thermalizing state.
The total imbalance $I(t)$ is plotted in Figs.~\ref{fig:evolution}(c) and \ref{fig:evolution}(d), which reveal more clearly the differences between two initial states. In general, for the thermalizing state, after about $30$ ns the population is only half photon in each qubit meaning there is no imbalance.

The distinct features of QMBS states can be further highlighted through the overlap between the product states and the eigenstates $|\langle \alpha  | E_n \rangle |^2$, which can be represented by the Fourier spectrum of the imbalance, as shown in Figs.~\ref{fig:evolution}(e) and \ref{fig:evolution}(f) for the QMBS and thermalizing states respectively. The peak value of the squared Fourier spectrum $g_\alpha(\omega)$ with the first-order domain eigenstates is $\omega_1 /2\pi \approx  21$ MHz. We test 120 random initial product states and find that the squared Fourier amplitudes $g_\alpha^2(\omega=\omega_1)$ of
QMBS states are unambiguously distinct from those of conventional
thermalizing states, as shown in Fig.~\ref{fig:evolution}(h).
The variations in the Fourier amplitude for rapidly thermalizing dynamics become suppressed as system size is increased.
Note that, for
the cases in Figs.~\ref{fig:evolution}(a-f), to carry out the exact simulations
is computationally impractical. To validate the experimental data numerically,
it is necessary to use a smaller system size, say $L=20$, whose results are
shown in the insets in Figs.~\ref{fig:evolution} (c-f), where the agreement
between numerical and experimental results is excellent.

The advantage of our experimental system -- the tunable effective couplings
between two nearest-neighbor qubits -- allows us to systematically probe the stability of QMBS states as the ratio of intra- and inter-dimer couplings
$J_\mathrm{a}/J_\mathrm{e}$ is varied. As shown in Fig.~\ref{fig:evolution}(g),
both the numerical and experimental results indicate that QMBS states emerge
consistently in the regime of $J_\mathrm{a}/J_\mathrm{e}>1$. Also, the finite
value of $g_\Pi(\omega_1)\approx 0.008$ for $J_\mathrm{a}/J_\mathrm{e}=1$
implies that the origin of the QMBS state is a many-body effect. 
In the regime of large coupling ($J_\mathrm{a,e}/2\pi>12$ MHz), the effective
Hamiltonian in Eq.~(\ref{eq:Hamiltonian}) is no longer accurate to describe
the system due to the population leakage to couplers.
Due to the fast growth of Hilbert space dimension in this case, we did not explore such large coupling regime.

To verify the persistence of the QMBS states for
different system sizes, we perform experimental measurements of chains of
sizes $L=12$, $14$, $16$, $20$, $24$, $28$, and $30$. The time evolution
of the imbalance, the entanglement entropy, and the four-qubit fidelity
were found to behave consistently for different system sizes, thereby
establishing the robustness of scarring in collective
states $|\Pi\rangle$ and $|\Pi^\prime\rangle$. The relatively small variations between the imbalance and the
entanglement entropy for different system sizes are due to the difference in
the cross couplings and the couplers. The Fourier amplitude
$g_\Pi(\omega_1)$ and the four-qubit fidelity $F_A(t_1)$ at the first
revival plateaus for $L > 16$, as shown in Fig.~\ref{fig:scaling}(a), whereas
the inverse of the Hilbert space dimension characterizing the scaling of a
random state shows an expected rapid exponential decrease with the system size.
The plateaued behavior in the scaling suggests that QMBS states
persist in the regime of large system size approaching the thermodynamic limit.

\section*{Many-body scars on a comb}

The programmable feature of our SC circuit allows us to emulate
topology beyond one dimension. We have experimentally studied QMBS states in a system with a more complex qubit structure illustrated in
Fig.~\ref{fig:comb}(a) -- a comb geometry which consists of a 1D ``backbone'' decorated with linear ``offshoots".  Previous studies of quantum comb systems with offshoots of random lengths were shown to exhibit localization, including ``compact" localized states for which  the localization length can vanish along the backbone~\cite{Maimaiti2017,HDGC:2020}. In our realization,  we take the offshoots to be of the same length, and we fix  the numbers of qubits and photons to be $L=20$ and $N=10$. We consider each offshoot to be a dimer and, as in the 1D case, we set the inter dimer couplings to  $J_\mathrm{e}/2\pi \simeq -6$ MHz and the intra dimer ones to $J_\mathrm{a}/2\pi \simeq -9$ MHz. In contrast to the chain geometry, the QMBS states in the comb geometry are
$|\Theta\rangle=|\mathrm{d}_+\mathrm{d}_+\cdots\rangle$ and
$|\Theta^\prime\rangle=|\mathrm{d}_-\mathrm{d}_-\cdots\rangle$. These states are also characteristically distinct from the conventional
thermalizing states, as revealed by the squared Fourier amplitude in
Fig.~\ref{fig:comb}(b). The striking contrast between a QMBS state and a
thermalizing state can be seen at a more detailed level from
Figs.~\ref{fig:comb}(c-d), which show the time evolution of the imbalance $I(t)$, four-qubit fidelity, and entanglement
entropy for state $|\Theta^\prime\rangle$ and a typical thermalizing state.

\section*{Discussion and Outlook}

In summary, we have experimentally realized QMBS states in a SC
circuit that emulates quantum many-body systems effectively described by a non-constrained spin-$1/2$ XY model with both one- and quasi-one-dimensional geometries. In contrast to existing experimental realizations in ultracold atomic systems~\cite{BSKL:2017,Su2022}, our work represents the first
experimental observation of QMBS states in a solid-state
device. Moreover, the underlying mechanism of scarring -- approximate decoupling of a hypercube subgraph of the Hilbert space in the computational basis -- is fundamentally distinct from other QMBS platforms.  Our study provides the first in-depth characterization of QMBS using quantum state tomography on large subsystems. By observing the population dynamics and entanglement entropy, we distinguished the weak ergodicity breaking associated with QMBS initial states from the conventional thermalizing states.

According to ``strong" ETH lore, large many-body systems with Hilbert dimension in the range of tens or hundreds of millions,  should exhibit ergodic dynamics. Our study indicates that QMBS states can arise in the SC processor even when the system size is this large, opening the door to investigating QMBS states and other many-body phenomena with an enormous Hilbert space in a feasible way, e.g.,
as in classical programmable computers. Our observation and characterization
of long-lived quantum states in complex and strongly interacting solid-state
systems with inevitable imperfections such as cross couplings between qubits, random disorders, and
environment-induced decoherence and dephasing lead to direct applications. For example, the robustness of QMBS states in solid-state systems can
substantially extend the coherence time of specific quantum information
operations such as the generation of GHZ states. Our work points to the need
to further investigate non-thermal states in experimental platforms to generate
QMBS states against quantum thermalization for quantum information and quantum
metrology applications~\cite{XZQL:2021}.


\section*{Acknowledgment}
The device was fabricated at the Micro-Nano Fabrication Center of Zhejiang University.
We acknowledge the support of the National Natural Science Foundation of
China (Grants No.~92065204, No.~U20A2076, No.~11725419, and No.~12174342), the National Basic
Research Program of China (Grants No.~2017YFA0304300), the Zhejiang Province Key Research and
Development Program (Grant No.~2020C01019), and the Key-Area Research and Development Program
of Guangdong Province (Grant No.~2020B0303030001). The work at Arizona State University is
supported by AFOSR through Grant No.~FA9550-21-1-0186.
Z.P. and J.Y.D. acknowledge support by EPSRC grants EP/R020612/1 and  EP/R513258/1, and by Leverhulme Trust Research Leadership Award RL-2019-015.

\section*{Author contributions}
 L.Y. proposed the idea. L.Y., Y.-C.L., J.Y.D. and Z.P. developed the theory and numerical simulation. P.Z., H.D. and Y.G. performed the experiment, and H.L. and J.C. fabricated the device supervised by H.W.. L.Z. and J.H. developed the measurement electronics. L.Y., H.W., Y.-C.L. and Z.P. co-wrote the manuscript. All authors contributed to the experimental setup, discussions of the results and development of the manuscript.

\section*{Competing interests}

The authors declare no competing financial interests.

\section*{Data and materials availability}

The data that support the findings
of this study are available from the corresponding authors
upon reasonable request.

%

\section*{Methods}

{\bf Device--}We use a superconducting quantum processor in a flip-chip package, which hosts
a square of $6\times6$ transmon qubits ($Q_i$) with $60$ couplers ($Q_c$), each
inserted in-between two neighboring qubits, as shown in
Fig.~\ref{fig:schematic}(a). Each qubit (coupler) is a quantum two-level system
with ground state $|0\rangle$ and excited state $|1\rangle$, whose energy
separation can be dynamically tuned in the frequency range $4.3-4.8$ GHz
($4.9-6.0$ GHz). Each qubit has individual microwave (XY) and flux (Z) controls
and it is capacitively coupled to a readout resonator for state discrimination.
Each coupler has an individual flux (Z) control and remains in the ground state
during the experiment. We use high-precision synchronized analog signals to
control the qubits and couplers, with microwave pulses for qubit XY rotations
and state readout, and square flux pulses for tuning the qubit and coupler
frequencies. A complete experimental sequence
consists of three stages: (1) state preparation where single-qubit $\pi$ pulses
are applied to half of the qubits, (2) multiqubit interaction stage where the
nearest neighboring qubit couplings are programmed by adjusting the couplers'
frequencies, and (3) the measurement stage where all qubits are jointly read
out. The values of the relevant qubit parameters such as the qubit operation frequencies,
energy relaxation times (with mean about $50$ $\mu$s) and single-qubit randomized
benchmarking fidelities (with mean about $0.993$) can be found in Table S1
of SM.

{\bf Effective model--}We derive the effective spin-$1/2$ XY model for our experimental superconducting processor.
The full Hamiltonian of the superconducting circuit-QED system with both qubits and couplers
is given by~\cite{NRKB:2018}
\begin{equation}
\begin{split}
	\mathcal{H}_\mathrm{full} /\hbar &= \sum^L_{i=1} \big( \omega_i\mathcal{S}^+_i\mathcal{S}^-_i
	+  \frac{\eta_i}{2}\mathcal{S}^+_i\mathcal{S}^+_i\mathcal{S}^-_i\mathcal{S}^-_i \big) \\
	&+ \sum^{L-1}_{c=1}
    \big( \omega_c\mathcal{S}^+_c\mathcal{S}^-_c + \frac{\eta_c}{2}\mathcal{S}^+_c\mathcal{S}^+_c\mathcal{S}^-_c\mathcal{S}^-_c \big) \\
	&+ \sum^{L}_{i,j=1} g_{ij} (\mathcal{S}^-_i \mathcal{S}^+_j + \mathcal{S}^-_{j} \mathcal{S}^+_{i})
	- \sum_{i,c} g_{ic}(\mathcal{S}^+_i\mathcal{S}^-_c+\mathcal{S}^-_i\mathcal{S}^+_c) ,
\end{split}
\label{seq:hamiltionian_exact}
\end{equation}
where $\omega_i\ (\omega_c)$ is the frequency of the $i$th qubit ($c$'s coupler), $\mathcal{S}^+_i\ (\mathcal{S}^-_i)$ is the creation (annihilation) operator of $Q_i$, $g_{ij}\ (g_{ic})$ is the coupling strength between $Q_i$ and $Q_j$ ($Q_c$), and the rotating wave approximation is imposed on the qubit-coupler and qubit-qubit couplings. The subscripts ``$i,j$'' and ``$c$'' represent the indices of qubits and couplers respectively. In experiments, the anharmonicity $\eta_i$ is much larger than the couplings between the nearest neighboring qubits (typically $\eta_i/g_{ij}> 50$), so the full Hamiltonian~(\ref{seq:hamiltionian_exact}) can be reduced to the spin-$1/2$ XY Hamiltonian:
\begin{equation}
\begin{split}
	\mathcal{H} /\hbar&= \sum^L_{i=1} \omega_i\mathcal{S}^+_i\mathcal{S}^-_i + \sum^{L-1}_{c=1} \omega_c\mathcal{S}^+_c\mathcal{S}^-_c \\
	&+ \sum^{L}_{i,j=1} g_{ij} (\mathcal{S}^-_i \mathcal{S}^+_j + \mathcal{S}^-_{j} \mathcal{S}^+_{i})
	- \sum_{i,c} g_{ic}(\mathcal{S}^+_i\mathcal{S}^-_c+\mathcal{S}^-_i\mathcal{S}^+_c).
\end{split}
\label{seq:hamiltionian}
\end{equation}
We apply the Schrieffer-Wolff transformation $\mathcal{U} = e^\mathcal{W}$ to the Hamiltonian with
\begin{displaymath}
\mathcal{W} = \sum_c \sum_{i}\frac{g_{ic}}{\Delta_{ic}}(\mathcal{S}^+_i\mathcal{S}^-_c-\mathcal{S}^-_i\mathcal{S}^+_c),
\end{displaymath}
since all qubits are far detuned from the couplers with $|\Delta_{ic}|=|\omega_i-\omega_c| \gg |g_{ic}|$. The effective Hamiltonian can then be approximated as
\begin{equation} \label{eq:Hamiltonian}
\mathcal{H}_\mathrm{eff} / \hbar \approx \sum_{i, j} J_{ij} (\mathcal{S}^-_i \mathcal{S}^+_j + \mathcal{S}^+_i \mathcal{S}^-_j) + \sum_{i} \Omega_{i} \mathcal{S}_i^+ \mathcal{S}_i^-   ,
\end{equation}
where the effective coupling strength and transition frequencies are given by
\begin{align}
	{J}_{ij} &= g_{ij}+   \sum_c g_{ic}g_{jc} \Big[ \frac{1}{\Delta_{ic }}+\frac{1}{\Delta_{jc}}    \Big], \\
	{\Omega}_{i} &= \omega_{i}+  \sum_c \frac{g^2_{ic}}{\Delta_{ic}},
\end{align}
respectively. The strength of the indirect coupling can be tuned by adjusting the coupler frequency, so the net coupling strength for $Q_i$ and $Q_j$ can be dynamically tuned over a wide range, typically from $-15\times2\pi$~MHz to $1\times2\pi$~MHz.

\begin{figure} [ht!]
\centering
\epsfig{figure=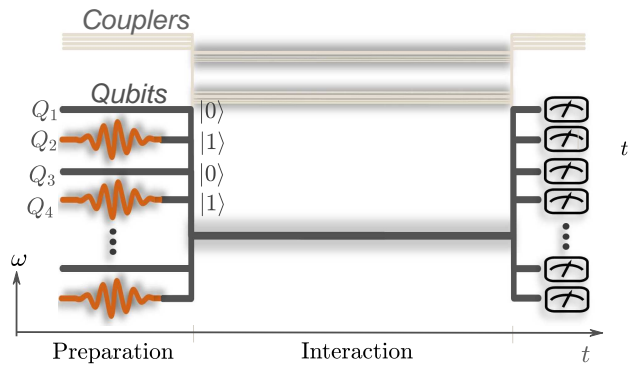,width=\linewidth}
\caption{ Experimental sequence with strongly interacting many-body dynamics, where
injecting a $\pi$ pulse (red wave pulse) serves to lift the two-level qubit
from the ground state to the excited state.     }
\label{fig:sequence}
\end{figure}

{\bf Experimental sequence--}Experimentally, we prepare a set of product states as initial states and
measure the final states of all qubits as a function of the interaction time
(see pulse sequence illustrated in Fig.~(\ref{fig:sequence}). A typical experimental session starts by preparing the initial product state
of all qubits: each qubit $Q_{i}$ is biased from its sweet spot to the
corresponding idle frequency, where we apply single-qubit $XY$ rotations. To
prepare a high-fidelity state, during this period the couplers are tuned such
that the net couplings between neighboring qubits are turned off. To
switch on the interactions among the qubits, we bias all qubits to the
interaction frequency and tune the coupler frequencies to turn on the couplings
between neighboring qubits. After the interaction process, we bias all qubits to their readout frequencies for measurements. All directly measured qubit occupation probabilities are corrected to eliminate the measurement errors.

\clearpage

\beginsupplement

\begin{center}
{\bf {\large Supplementary Materials}}
\end{center}


\maketitle

\section{Experimental Details}

{\bf Device information.}
The superconducting quantum processor is composed of $6\times6$ transmon qubits with tunable couplers between adjacent qubits in rows and columns. Each qubit has individual microwave (XY) and flux (Z) controls, and is capacitively coupled to its corresponding readout resonator for state discrimination. Each coupler has an individual flux (Z) control and remains in the ground state during an experiment.
All qubits and couplers are located on a sapphire substrate (top chip), and most portion of the control/readout lines and readout resonators are located on a silicon substrate (bottom chip). Both chips have lithographically defined base wirings made of tantalum, and are galvanically connected via 9 \text{$\mu m$}-tall indium bumps directly deposited on tantalum. The indium bumps are not only for ground connectivity, but also for passing through control signals from the bottom chip, which connects to external circuitry via aluminum bonding wires, to the top chip where the qubits and couplers are located.

Experiments are carried out using systems of up to $30$ transmon qubits and $29$ transmon-type couplers. Parameters benchmarking the single-qubit performances are listed in Table~\ref{tab:dev_para}. The couplers are used to tune the coupling strength between two neighboring qubits, where the tunable coupling value is characterized by the pairwise one-photon swapping dynamics. Figure~\ref{sfig:effective_coupling} shows an example of adjusting this coupling by tuning the coupler's frequency.

We also measure all dominant cross couplings $J_\mathrm{x}$ between qubits with separation distance $\sqrt{2}a_0$. Most values of $J_\mathrm{x}/2\pi$ fall in the range $[0.3,1.2]$ MHz, as shown in Fig.~\ref{sfig:Jx}.


\section{Numerical prediction of QMBS in small systems}

Quantum thermalization in isolated many-body systems is associated with the properties of their energy eigenstates. If the system obeys the Eigenstate Thermalization Hypothesis (ETH), all eigenstates away from the edges of the spectrum will behave similarly to thermal ensembles. For example, the expection value of any generic observable with respect to an eigenstate will only depend on its energy, and will match the prediction of the micro-canonical ensemble in the thermodynamic limit. As a consequence, quenching from a generic quantum state will lead to fast thermalization of this observable.
Integrable systems such as the 1D spin-$1/2$ XY model or the SSH chain strongly violates the ETH due to their extensive number of conserved quantities.
However, integrability is very fragile and will be broken in our SC device due to the presence of uneven cross couplings between non-nearest-neighbor spins, even if these are much smaller in strength that nearest-neighbor couplings.
Here we numerically demonstrate that this leads to thermalization of the system for the overwhelming majority of states. However, a few special eigenstates are still violating the ETH, and this causes non-ergodic dynamics from a small set of basis states.
We present numerical predictions of these QMBS states in systems of chain topology with regular ($J_\mathrm{nn}$) and irregular ($J_\mathrm{x}$) non nearest-neighbor couplings as well as a comb tensor system with irregular couplings $J_\mathrm{x}$.

{\bf Quantum fidelity for scarred dynamics.} The dynamical evolution of a quantum state can be written as $|\Psi(t)\rangle=\sum_n\langle E_n | \Psi(0) \rangle e^{-i E_n t}|E_n\rangle$, where $E_n$ and $|E_n\rangle=\sum_\alpha c_{n,\alpha}|\alpha\rangle$ are the eigenenergy and eigenvector of the $n$-th eigenstate, respectively, and $|\alpha\rangle$ labels the computational basis states.

In a conventional thermalizing system, for any physical initial state (e.g., chosen to be one of the product states $|\alpha\rangle$),  its squared overlap with the energy eigenstates $|\langle E_n|\Psi(0)\rangle|^2$ is homogeneously distributed. Time evolution of the fidelity, defined as ${F}(t)=|\langle \Psi(0)|\Psi(t)\rangle|^2=\sum_n |\langle \Psi|E_n\rangle |^2 e^{-iE_n t}$, will rapidly decay to zero as the phases are essentially random, with tiny fluctuations proportional to the inverse Hilbert space dimension. This means that the initial information encoded in $|\Psi(0)\rangle$ will uniformly diffuse across the entire Hilbert space and, at late times, it cannot be measured by any local operator.


On the contrary, for the slowly thermalizing QMBS state, the squared overlap $|\langle E_n | \Psi (0) \rangle |^2$ is appreciable only for a few special scarred eigenstates which are effectively evenly spaced in energy. In this case, the time evolution of the fidelity can be approximated by only taking into account these states as $F(t) \approx \sum_{s_k} |\langle E_{s_k}|\Psi(0)\rangle |^2 e^{-iE_{s_k} t}$, where $E_{s_k}$ denotes the eigenenergy of the $k$th scarred eigenstate. Due to the approximately equal energy spacing $\Delta E$, the phases will be coherent again after a time $t=2\pi/\Delta E $, leading to revivals of the wavefunction.

\begin{figure} [tb]
\centering
\includegraphics[width=\linewidth]{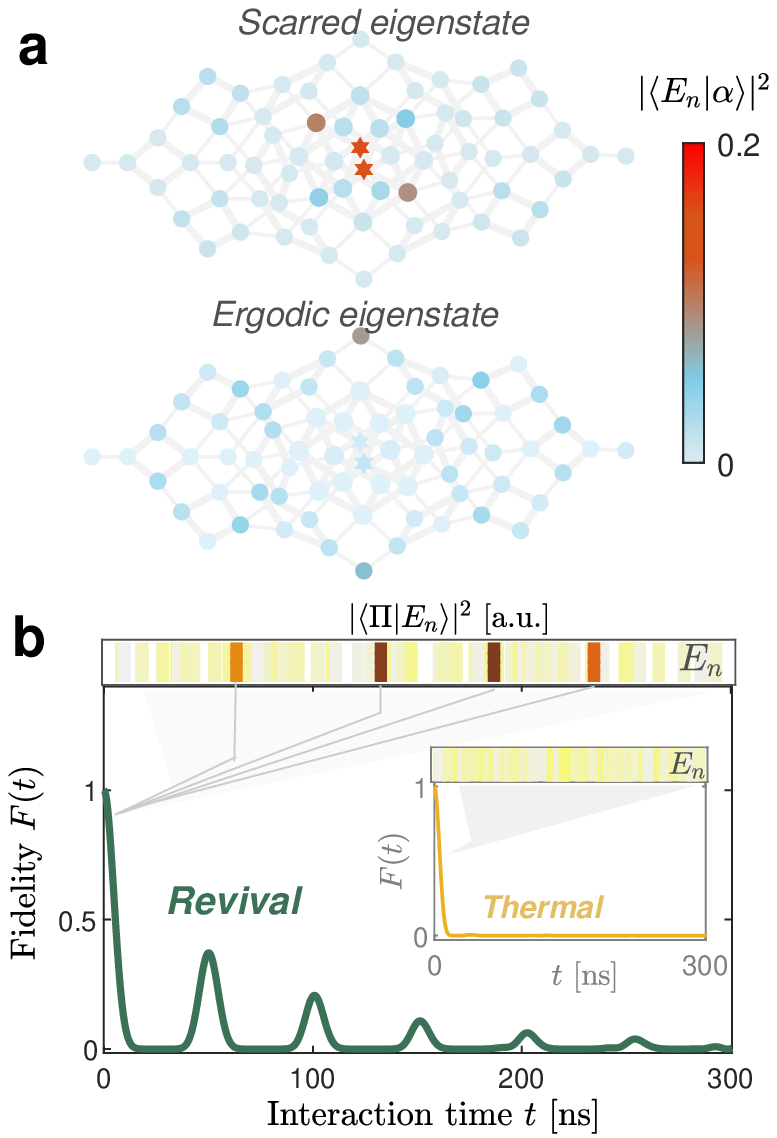}
\caption{
(a) Visualizing a selected scarred eigenstate and an ergodic eigenstate in the computational basis. Each vertex represents a computational basis state $|\alpha\rangle$, with the color denoting its contribution to the given eigenstate, $|\langle \alpha | E\rangle|^2$. One observes that the QMBS state is strongly concentrated on a few basis states, with the highest overlap on the two QMBS states $|\Pi\rangle$ and $|\Pi^\prime \rangle$ indicated by hexagrams. In contrast, the ergodic eigenstate is uniformly spread across all basis states.
The eigenstates are obtained for the chain geometry with photon number $N=L/2=4$ and coupling ratio $J_\mathrm{a}/J_\mathrm{e}=1.5$. Light gray line segments represent the various inter-qubit couplings $J_{ij}$.
(b) Simulated time evolution of the fidelity of the QMBS state $|\Pi\rangle$ for $N=L/2=7$, where the thermalization process is much slower than that of a random initial basis state (inset).  Top row: squared overlap between the initial product state and the eigenstates $|\langle \alpha | E_n \rangle|^2$. Coupling parameter values are $J_\mathrm{a}/2\pi=-9$, $J_\mathrm{e}/2\pi=-6$, and $J_\mathrm{x}/2\pi\in[0.3,1.2]$ MHz.
}
\label{fig:fidelity}
\end{figure}

To illustrate these points, in Fig.~\ref{fig:fidelity} we present exact diagonalization study of scarred eigenstates and fidelity dynamics, contrasting them with thermalizing states.
We consider a linear array of $L$ qubits at half filling, $N=L/2$, with the intra- and inter-qubit couplings in ratio $J_a/J_e=1.5$. We also include next-nearest neighbor couplings $J_x$ which lead to the thermalization of the system. The support of energy eigenstates on product states, shown in Fig.~\ref{fig:fidelity}(a), reveals a concentration of a QMBS eigenstate on few computational basis states. In contrast, a typical thermalizing state is uniformly spread across the Hilbert space. Moreover, as explained in the main text, the QMBS eigenstates can be probed by quenching the system from the initial state $|\Pi\rangle$, which is a product state of dimers alternating between $01$ and $10$ local configurations. For such an initial state, quench dynamics leads to revivals in the quantum fidelity [Fig.~\ref{fig:fidelity}(b)] which are absent for thermalizing initial states [inset of Fig.~\ref{fig:fidelity}(b)].

{\bf QMBS states in a chain system subject to regular perturbations.}
As cross couplings in our SC device are irregular in strength, in our numerical simulations we choose them to be random in value in an experimentally--relevant range. However, to show that there is no need to add randomness to break integrability and see the appearance of scarring, we will first ignore the cross coupling and a regular longer-range perturbation instead.
We thus apply the next-next-nearest-neighbor couplings $J_\mathrm{nn}=J_{i,i+3}$ that preserve the reflection symmetry. We consider the chain structure of $L=18$ and $N=9$ with open boundary conditions for the following parameter values: $J_\mathrm{a}/2\pi=1.5J_\mathrm{a}/2\pi=-6$ and $J_\mathrm{nn}/2\pi=0.6$ MHz. The collective dimerized states $|\Pi\rangle$ and $|\Pi^\prime \rangle$ specified in Fig.~1(b) of the main text have a large overlap with a set of special eigenstates marked by red hexagrams in Figs.~\ref{sfig:scar_ssh_Jnn_pbc}(a) and \ref{sfig:scar_ssh_Jnn_pbc}(b). These eigenstates possess lower entanglement entropy than the conventional states whose entropy obeys the volume law. The time evolution of the fidelity, imbalance, and entanglement entropy of the state $|\Pi\rangle$ and other randomly chosen initial product states are also shown in Figs.~\ref{sfig:scar_ssh_Jnn_pbc}(c-e).
It can be seen that the state $|\Pi\rangle$ is dynamically long-lived in comparison with other initial states.

On top of the $J_\mathrm{nn}$ couplings, one can also add some perturbations on the frequency term $\Omega_i$. These can take the form of impurities located at the end of the chain $\Omega_{L}/2\pi=\Omega_{L-1}/2\pi=3$ MHZ, or of a staircase potential $\Omega_{2n-1}/2\pi=\Omega_{2n}/2\pi=0.8n$ MHz with $n=1,2,\cdots,N$.
In both cases, the perturbation is constant within each dimer and we expect it to leave the $|\Pi \rangle$ state and all dimerized states unaffected. However, because of the difference between dimers, it should lead to faster thermalization in the rest of the Hilbert space. This is confirmed
by our numerical simulation, as shown in Fig.~\ref{sfig:ssh_Jnn_Om}.

{\bf QMBS states in a chain subject to irregular perturbations.}
In the experimental superconducting circuit in Fig.~1(a) of the main text,  cross couplings $J_\mathrm{x}$ are the ones that break the reflection symmetry as well as the integrability of the SSH chain. Due to their variability of their strengths, we consider these to be random in value within the range $2\pi\times[0.3,1.2]$ MHz, which matches what we measure in the actual experiment. We investigate the QMBS states in chain systems with the SSH topology based on the effective Hamiltonian -- Eq.~(\ref{eq:Hamiltonian}) in the main text -- with the
experimental parameter values $J_\mathrm{a}/2\pi=J_\mathrm{e}/2\pi=-6$ MHz.
The numerical results for $L=18$ and $N=9$ are summarized in Fig.~\ref{sfig:scar_ssh}. The observed Wigner-Dyson distribution of the level spacing indicates the ``quantum chaotic'' nature of the system with a Gaussian
distribution of the density of states (DOS), as shown in Fig.~\ref{sfig:scar_ssh}(a). The values of the entanglement entropy of all the eigenstates is consistent with a volume law, as it approaches the approximate Page value $L/2\mathrm{ln}2-1/2$ near the middle of the spectrum, as shown in Fig.~\ref{sfig:scar_ssh}(b). The squared overlap between the collective dimerized state $|\Pi\rangle$ and the eigenstates $|E_n\rangle$ is shown in Fig.~\ref{sfig:scar_ssh}(c). On top of the bulk of thermal states, several towers of states with a higher overlap with the $|\Pi\rangle$ states are clearly visible. These are approximately equally spaced in energy, leading to the oscillatory dynamics that can be seen in the fidelity and imbalance when starting from the $|\Pi\rangle$ state, as shown in Figs.~\ref{sfig:scar_ssh}(d) and \ref{sfig:scar_ssh}(e) respectively. The entanglement entropy of the scarred product state increases with time more slowly than for conventional states, as shown in Fig.~\ref{sfig:scar_ssh}(f). Figure~\ref{sfig:scar_ssh}(g) shows the scaling of the fidelity density of the scarred state, defined as $(1/L)\mathrm{ln} F(t=t_1)$, where $t_1$ is the time of the first fidelity revival. The fidelity density converges to a constant value of about $-0.09$, which is much higher than the scaling of $1/D$ (with $D$ the dimension of the Hilbert space at half-filling) expected from random states. In addition, the revival behavior of the fidelity can be enhanced by increasing the coupling ratio $J_\mathrm{a}/J_\mathrm{e}$.
For example, for $J_\mathrm{a}/J_\mathrm{e}=2$ and $2.5$, the peak values of the fidelity and imbalance become much larger than that for
$J_\mathrm{a}/J_\mathrm{e}=1.5$. Simultaneously, the entanglement entropy is smaller, as shown by the light green curves in Fig.~\ref{sfig:scar_ssh}(f).

{\bf QMBS states in a comb tensor system subject to irregular perturbations.}
The comb tensor system has a quasi-one-dimensional topology. Figure~\ref{sfig:scar_comb} shows the simulation results for $J_\mathrm{a}/2\pi=1.5 J_\mathrm{e}/2\pi=-9$ MHz and $J_\mathrm{x}/2\pi\in[0.3,1.2]$ MHz, which are similar to those of the 1D system. A difference is that the collective dimer states in the comb tensor system are $|\Theta \rangle=|101010\cdots\rangle$ and $| \Theta^\prime \rangle=|010101\cdots\rangle$. Such a system is ergodic even without cross coupling and the level-spacing distribution in each symmetry sector is of the Wigner-Dyson type.


\section{Population Dynamics Associated with Scarring States}

The revival fidelity over time can be expressed in term of the overlap between the initial state and the eigenstates of the system as,
\begin{equation}
F(t)=|\langle\psi(t)|\psi(0)\rangle|^2 = \sum_n  |c_{\psi n}|^2 e^{-i E_n t},
\end{equation}
where $c_{\psi n} =\langle \psi(0)|E_n\rangle $ and the time evolution of the wavefunction is given by
\begin{displaymath}
|\psi(t)\rangle = \sum_n c_{\psi,n} e^{-i E_n t} |E_n\rangle.
\end{displaymath}

Because of the difficulty in directly measuring the fidelity, we use an experimentally feasible quantity known as the imbalance $I(t)$ to replace the fidelity, which is defined as
\begin{equation}
I(t) = \frac{1}{L}\sum^L_{i=1}  \langle \mathcal{S}^z_i(0)\rangle \langle \mathcal{S}^z_i(t) \rangle,
\end{equation}
where $\langle \mathcal{S}_i^z (t) \rangle=2n_i(t)-1$ with $n_i(t)$ being the population of $Q_i$. Here, we consider for the initial states the set of basis states ${|\alpha \rangle}$, where $\alpha=1,2,\cdots,D$ with $D$ being the dimension of the Hilbert space. The $n$-th eigenstate can be written in this basis as $|E_n\rangle = c_{n \alpha} | \alpha \rangle$. The time evolution of the initial product state is determined by $|\alpha(t)\rangle = \mathcal{U}(t)|\alpha\rangle$, where
\begin{displaymath}
\mathcal{U}(t)=e^{-i\mathcal{H}t}=\sum_n e^{-i E_n t}|E_n\rangle\langle E_n|.
\end{displaymath}
We get
\begin{equation}
\begin{split}
	\langle  \mathcal{S}_i^z (t) \rangle &=\langle\alpha(t) |   \mathcal{S}_i^z  | \alpha (t) \rangle   \\
	&=\sum_{n}\sum_m c^\ast_{\alpha n}c_{\alpha m}e^{-i(E_m-E_{n})t} \langle E_{n}  |  \mathcal{S}_i^z    |E_m\rangle\\
	&=\sum_n\sum_m   c^\ast_{\alpha n}   c_{\alpha m}e^{-i(E_m-E_n)t}
	\sum_{\beta}\sum_\gamma c^\ast_{\gamma n}c_{\beta m}\langle\gamma|  \mathcal{S}_i^z | \beta \rangle  \\
    &=\sum_n\sum_m \sum_{\beta}  s_{\beta,i}  c^\ast_{\alpha n}c_{\alpha m}  c^\ast_{\beta  n}   c_{\beta m}e^{-i(E_m-E_n)t},
\end{split}
\end{equation}
where $s_{\beta,i}=1$ or $-1$ for states $|1 \rangle_i$ or $| 0 \rangle_i$. The imbalance is thus given by
\begin{equation}
\begin{split}
I(t)=\frac{1}{L} \sum_{i=1}^L \sum_n\sum_m \sum_{\beta}  s_{\alpha,i}  s_{\beta,i} \left(  c^\ast_{\alpha  n}c_{\alpha m}  c^\ast_{\beta  n} c_{\beta m} \right) e^{-i(E_m-E_n)t},
\end{split}
\end{equation}
where the subscripts $i,j$, $n,m$, and $\alpha,\beta,\gamma$ are the qubit, eigenstate and basis state indices, respectively.

The imbalance dynamics of the scarred states is dominated by a set of specific eigenstates, such as those indicated by the red crosses in Fig.~\ref{sfig:scar_ssh}(d). As shown in Figs.~\ref{sfig:scar_ssh}(c) and \ref{sfig:scar_ssh}(d), the squared overlap of the eigenstates with the scarred state $|\Pi\rangle$ exhibits $L/2+1$ peaks with a constant energy interval $\Delta E$, while a conventional thermalizing state has a uniform overlap with the eigenstates. For example, for the system of size $L=14$ with $8$ specific eigenstates, the squared overlaps denoted as $a^2_{  1}$, $a^2_{ 2}$, $a^2_{ 3}$, and $a^2_{ 4}$ with the respective eigenenergies $\pm \Delta/2$, $\pm 3\Delta/2$, $\pm 5\Delta/2$, and $\pm 7\Delta/2$ satisfy the inequality $a^2_{\pm 1} >a^2_{\pm 2}>a^2_{\pm 3}>\cdots$.

The main terms in the overlap can be estimated from the formula of imbalance that can be written as
\begin{equation}
I(t) \approx I_0 +  \sum_{k,l=-4}^4   a^2_k a^2_l    e^{-i (k-l)\Delta E t},
\end{equation}
where $I_0$ is the contribution from other eigenstates. For $a^4_{\pm 1} \gg a^4_{\pm 2}\gg  a^4_{\pm 3}\gg  a^4_{\pm 4} $ and $a^2_{k} = a^2_{-k}$ (with $k=1,2,3,4$), we have
\begin{equation}
\begin{split}
	I(t)-I_0 & \approx 2(2a^2_{4} a^2_{3}+2a^2_{3} a^2_{2}+2a^2_{2}a^2_{1}+a^4_{1} ) \cos(\Delta Et) \\
		& + 2(2a^{2}_1a^2_{2}+2a^2_{3}a^2_{1}+2a^2_{2} a^2_{4})\cos(2\Delta Et)  \\
        & + 2(2a^2_{4}a^2_{1}+2a^2_{3}a^2_{1}+a^4_{2})\cos(3\Delta Et) \\
		& + 2(2a^2_{4}a^2_{1}+2a^2_{3}a^2_{2})\cos(4\Delta Et)  \\
        & + 2(2a^2_{4}a^2_{2}+a^4_{3})\cos(5\Delta Et) \\
		& + 4a^2_{4}a^2_{3} \cos(6\Delta Et)+2a^4_{4} \cos(7\Delta Et) \\
        & \approx  (4a^2_{2}a^2_{1}+2 a^4_{1}) \cos(\Delta Et).
\end{split}
\end{equation}
This example shows that the Fourier transformation of the imbalance dynamics contains only one peak, as shown in Fig.~2 of the main text, whereas that of the fidelity dynamics has a set of peaks. The conventional thermalized states are uniformly distributed over the eigenstates, so their imbalance dynamics rapidly decay to zero.

\section{Comparison between Experimental and Numerical Results in Systems of Varying Sizes}

{\bf Population dynamics at each qubit.}
Figure~\ref{sfig:eachQ_ni}(a) shows the experimentally measured and numerically calculated imbalances versus the interaction time for a chain system of size $L=20$, which agree with each other. The squared Fourier amplitudes of the scarred states $|\Pi\rangle$ ($|\Pi^\prime\rangle$) are much higher than those of the conventional thermalizing states, as shown in Fig.~\ref{sfig:eachQ_ni}(b). The experimental and numerical population evolution of each qubit in a $20$-qubit chain for $J_\mathrm{a}/J_\mathrm{e}=1.5$ are shown in Fig.~\ref{sfig:eachQ_ni}(c).

{\bf Size dependence of the characterizing quantities.}
We measure the time evolution of the imbalance, fidelity, and four-qubit entanglement entropy for different system sizes, as shown in Fig.~\ref{sfig:scaling}. For the collective dimerized states, these characterizing quantities exhibit highly consistent behaviors for sizes ranging from $L=12$ to $30$. Note that the data here has been used to generate the synthesized results in Fig.~4 of the main text.

\section{QMBS states with periodic boundary conditions}
We find scarred features of the collective dimerized states become weak with the periodic boundary condition (PBC) for $L/2\in$ odd, which is confirmed in the experment of $30$-qubit chain with a periodic boundary condition, as shown as the fidelity and entanglement entropy in Fig.~\ref{sfig:pbc}.


\begin{table*} [ht!]
    \centering
    \addtolength{\tabcolsep}{+2pt}
\caption{\small Device parameters: $\omega^{0}_j$ is the transition frequency of $Q_j$ with zero flux bias, known as the sweet spot; $\omega^{\text{i}}_j$ is the idle frequency of $Q_j$ where single-qubit XY rotations are applied and the average single-qubit gate error $e_{sq}$ is measured via randomized benchmarking. All qubits are flux (Z) biased to $\omega^{\text{I}}/2\pi \approx 4.375$~GHz to activate the interaction, where the energy relaxation time $T_{1, j}$ and the Ramsey dephasing time $T^*_{2, j}$ of each qubit $Q_j$ are measured.
}
\begin{adjustbox}{width=4in}
\begin{tabular}{l|c|cc|ccc}
        \hline
        \hline
        \quad & $\omega^{0}_j / 2\pi$ (GHz) & $\omega^{\text{i}}_j / 2\pi$ (GHz) & $e_{sq}$ ($\%$) & $T_{1, j}$ ($\mu$s) & $T^*_{2, j}$ ($\mu$s) \\
        \hline
        $Q_{1 }$&   4.826&     4.795&    0.26&      71.5&     2.2\\
        $Q_{2 }$&   4.880&     4.420&    0.18&      75.5&     2.3\\
        $Q_{3 }$&   5.025&     4.370&    0.62&      78.0&     2.0\\
        $Q_{4 }$&   4.984&     4.310&    0.69&      62.5&     1.8\\
        $Q_{5 }$&   4.906&     4.285&    0.61&      53.3&     1.9\\
        $Q_{6 }$&   4.936&     4.810&    0.58&      37.1&     2.9\\
        $Q_{7 }$&   4.963&     4.375&    0.80&      70.7&     1.9\\
        $Q_{8 }$&   4.949&     4.305&    0.69&      43.5&     2.4\\
        $Q_{9 }$&   4.992&     4.695&    0.60&      56.8&     2.0\\
        $Q_{10}$&   4.856&     4.745&    0.40&      60.2&     2.0\\
        $Q_{11}$&   4.855&     4.360&    0.70&      63.4&     2.0\\
        $Q_{12}$&   4.825&     4.490&    0.48&      60.6&     2.4\\
        $Q_{13}$&   4.875&     4.772&    0.55&      55.9&     2.4\\
        $Q_{14}$&   4.845&     4.720&    0.70&      48.6&     2.3\\
        $Q_{15}$&   4.904&     4.350&    1.94&      48.4&     1.2\\
        $Q_{16}$&   5.043&     4.320&    0.74&      49.9&     2.0\\
        $Q_{17}$&   5.012&     4.660&    1.31&      53.8&     2.2\\
        $Q_{18}$&   5.025&     4.758&    0.43&      55.9&     1.2\\
        $Q_{19}$&   4.984&     4.436&    0.73&      35.6&     1.2\\
        $Q_{20}$&   4.975&     4.800&    0.49&      34.9&     1.7\\
        $Q_{21}$&   4.976&     4.655&    0.48&      35.7&     1.3\\
        $Q_{22}$&   4.927&     4.353&    0.71&      45.9&     1.5\\
        $Q_{23}$&   4.947&     4.749&    0.74&      49.7&     1.8\\
        $Q_{24}$&   4.975&     4.625&    0.73&      52.2&     2.0\\
        $Q_{25}$&   4.974&     4.801&    0.76&      37.0&     1.7\\
        $Q_{26}$&   4.916&     4.340&    0.92&      50.1&     1.8\\
        $Q_{27}$&   4.890&     4.430&    0.56&      50.9&     1.2\\
        $Q_{28}$&   4.920&     4.640&    0.72&      47.9&     1.3\\
        $Q_{29}$&   4.874&     4.710&    0.40&      46.2&     1.5\\
        $Q_{30}$&   4.844&     4.330&    0.40&      70.3&     2.5\\
        \hline
        Average &       -&         -&    0.66&      53.4&     1.9\\
        \hline
\end{tabular}
 \end{adjustbox}
\label{tab:dev_para}
\end{table*}

\begin{figure*} [ht!]
\centering
\epsfig{figure=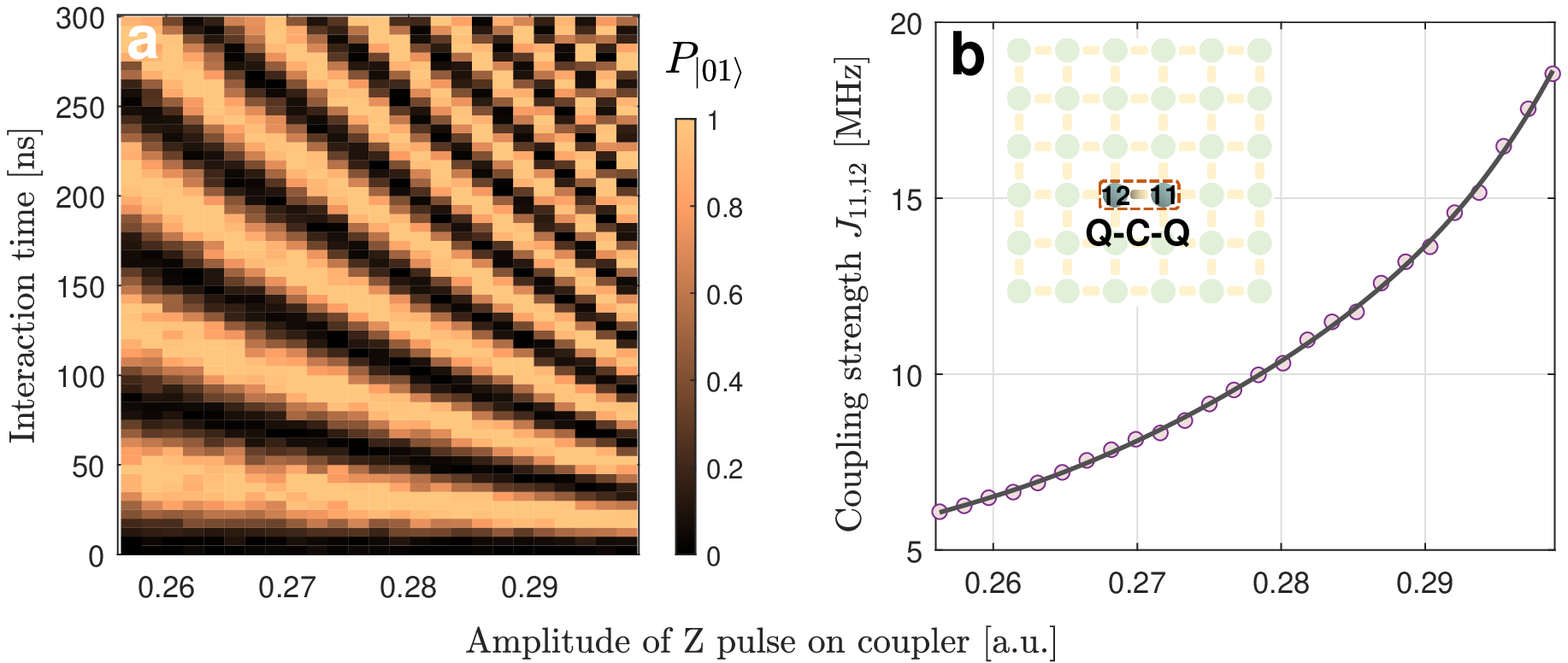,width=6in}
\caption{Effective coupling between two nearest-neighbor qubits as a function of the amplitude of the flux (Z) bias on the
coupler. We start by preparing $Q_{11}$-$Q_{12}$ in $|10\rangle$, then turn on the effective coupling by applying the flux
(Z) bias to position the coupler at $\sim5$~GHz (originally idling at $\sim6$~GHz) for the interaction time, and finally
measure the resulting probability $P_{|01\rangle}$ for the two qubits to be in the state $|01\rangle$. (a) $P_{|01\rangle}$ as a function
of both the Z bias amplitude and the interaction time. (b) Effective coupling strength between $Q_{11}$ and $Q_{12}$ versus the Z bias amplitude, obtained by Fourier transform of the data in (a). The schematic inset displays a
qubit-coupler-qubit structure during the measurement.}
\label{sfig:effective_coupling}
\end{figure*}

\begin{figure*} [ht!]
\centering
\epsfig{figure=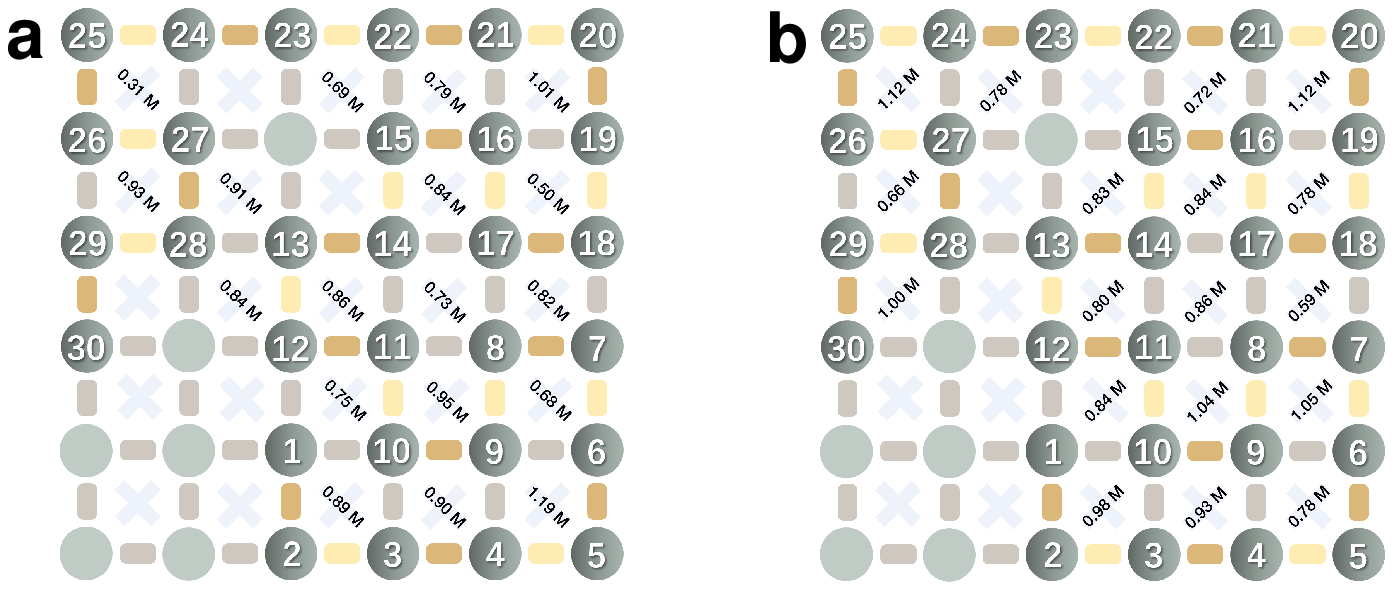,width=6.5in}
\caption{The dominant cross coupling $J_\mathrm{x}/2\pi$ [Hz] value from experimental measurements.  }
\label{sfig:Jx}
\end{figure*}

\begin{figure*} [ht!]
\centering
\epsfig{figure=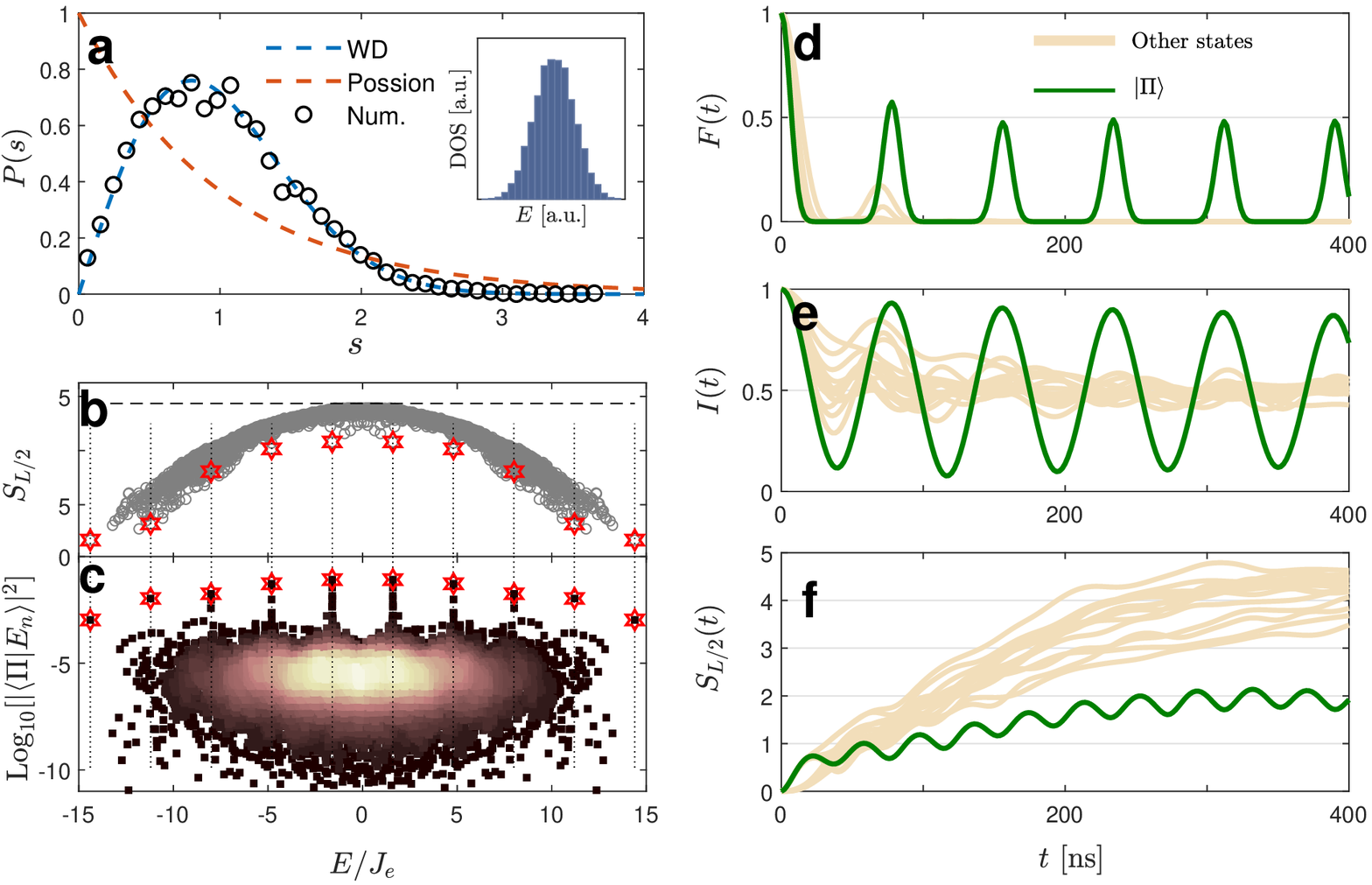,width=0.9\linewidth}
\caption{Example of QMBS in a chain. The chain parameters are $L=18$, $N=9$, $J_\mathrm{a}/2\pi=-6$, $J_\mathrm{e}/2\pi=-4$,
and $J_\mathrm{nn}/2\pi=0.72$ MHz. (a) Level-spacing statistics after resolving the inversion symmetry. The black circles show the numerical data, while the dotted lines show the expected distributions for an integrable (Poisson) and a chaotic (Wigner-Dyson) model. The inset shows the density of states.
(b) Entanglement entropy of all eigenstates, with the horizontal dashed line
showing the approximate Page value $(L/2)\mathrm{ln}2-1/2$. (c) Squared overlap between the collective dimerized state $|\Pi\rangle$ and
the eigenstates $|E_n\rangle$. In (b) and (c), the red hexagrams mark the major eigenstates that contribute to the scarring
dynamics. (d-f) Time evolution of fidelity, imbalance, and entanglement entropy for the collective dimerized state (green)
and other random states (light orange).}
\label{sfig:scar_ssh_Jnn_pbc}
\end{figure*}

\begin{figure*} [ht!]
\centering
\epsfig{figure=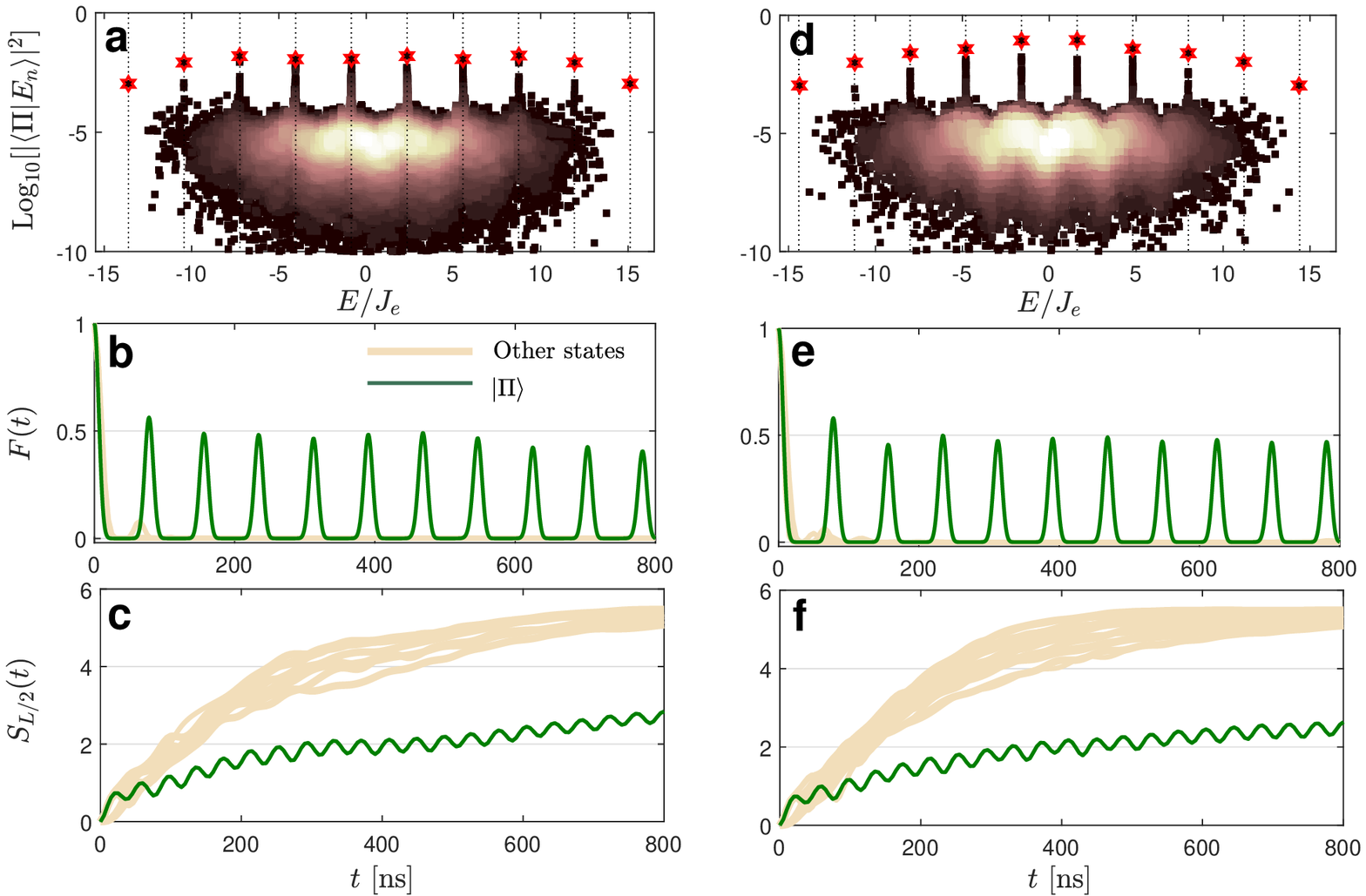,width=6.5in}
\caption{Example of QMBS in a chain. The chain parameters are $L=18$, $N=9$, $J_\mathrm{a}/2\pi=-6$, and  $J_\mathrm{e}/2\pi=-4$ MHz. (a-c) Squared overlap between the collective dimerized state $| \Pi \rangle$ and the eigenstates $|E_n\rangle$, time evolution of fidelity, and entanglement entropy for the collective dimerized state (green) and other randomly chosen product states (light orange) for
  $J_\mathrm{nn}/2\pi=0.7$ MHz and impurities $\Omega_{L}/2\pi=\Omega_{L-1}/2\pi=3$ MHz. (d-f) Similar
  to (a-c), but with a staircase perturbation
  $\Omega_{2n-1}/2\pi=\Omega_{2n}/2\pi=0.8n$ MHz with $n=1,2,\cdots,N$.}
\label{sfig:ssh_Jnn_Om}
\end{figure*}

\begin{figure*} [ht!]
\centering
\epsfig{figure=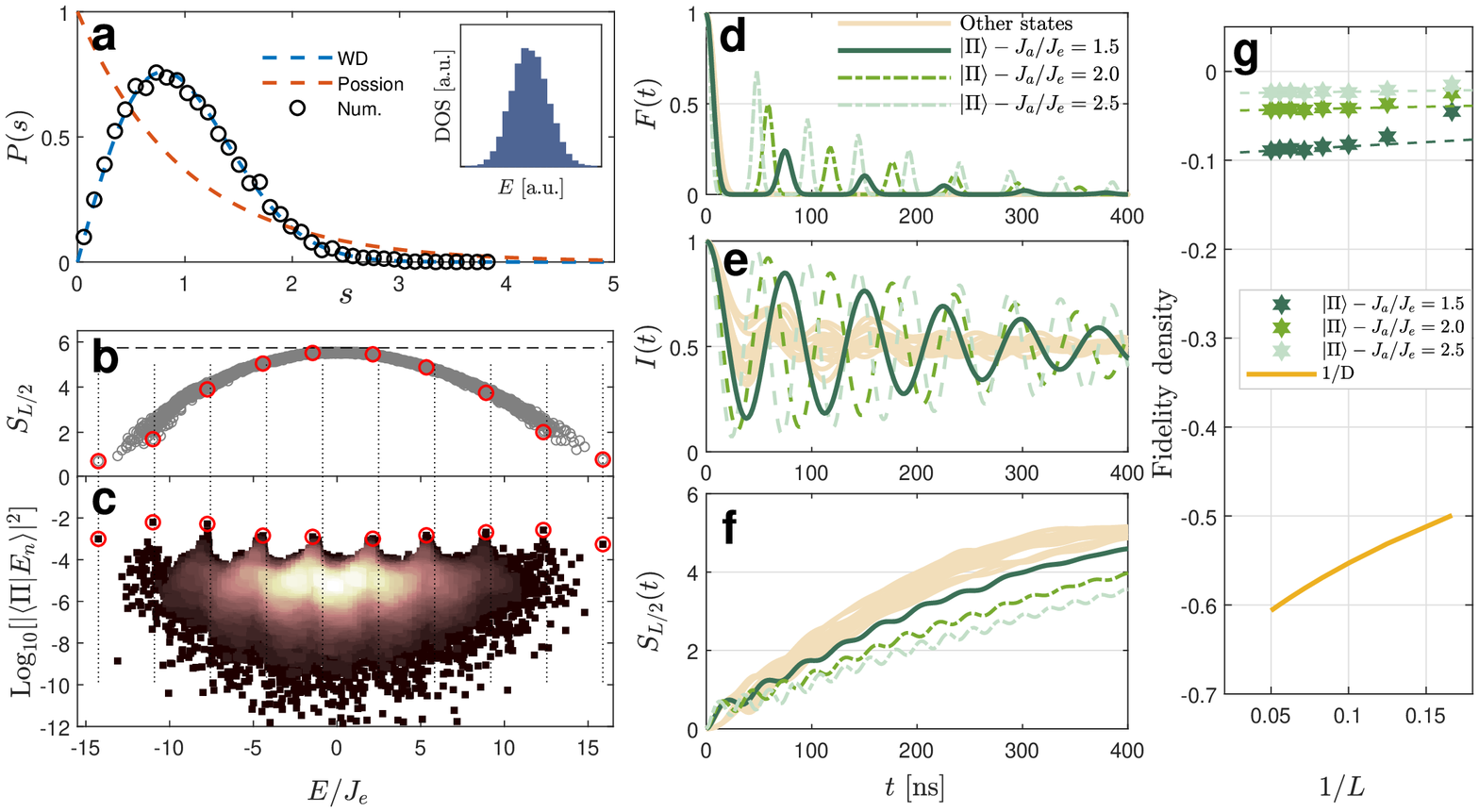,width=\linewidth}
\caption{QMBS with random cross couplings. The chain parameters are $L=18$, $N=9$, $J_\mathrm{a}/2\pi=-6$, $J_\mathrm{e}/2\pi=-4$,
and $J_\mathrm{x}/2\pi\in[0.3,1.2]$ MHz. (a) Level-spacing statistics, the black circles show the numerical data, while the dotted lines show the expected distributions for an integrable (Poisson) and a chaotic (Wigner-Dyson) model. The inset shows the density of states.
(b) Entanglement entropy of all eigenstates, with the horizontal dashed line
showing the approximate Page value $(L/2)\mathrm{ln}2-1/2$. (c) Squared overlap between the collective dimerized state $|\Pi\rangle$ and
the eigenstates $|E_n\rangle$. In (b) and (c), the red hexagrams mark the major eigenstates that contribute to the scarring
dynamics. (d-f) Time evolution of fidelity, imbalance, and entanglement entropy for the collective dimerized state (green)
and other random states (light orange). (g) Fidelity density of the QMBS state $|\Pi\rangle$ and inverse Hilbert space dimension $1/D$ as a function of $1/L$. For the cases with coupling ratio $J_\mathrm{a}/J_\mathrm{e}=2$ and $2.5$, $J_\mathrm{e}/2\pi=-4$
MHz and $J_\mathrm{x}/2\pi\in[0.3,1.2]$ MHz are fixed and $J_\mathrm{a}/2\pi$ is increased to $-8$ and $-10$ MHz respectively.}
\label{sfig:scar_ssh}
\end{figure*}

\begin{figure*} [ht!]
\centering
\epsfig{figure=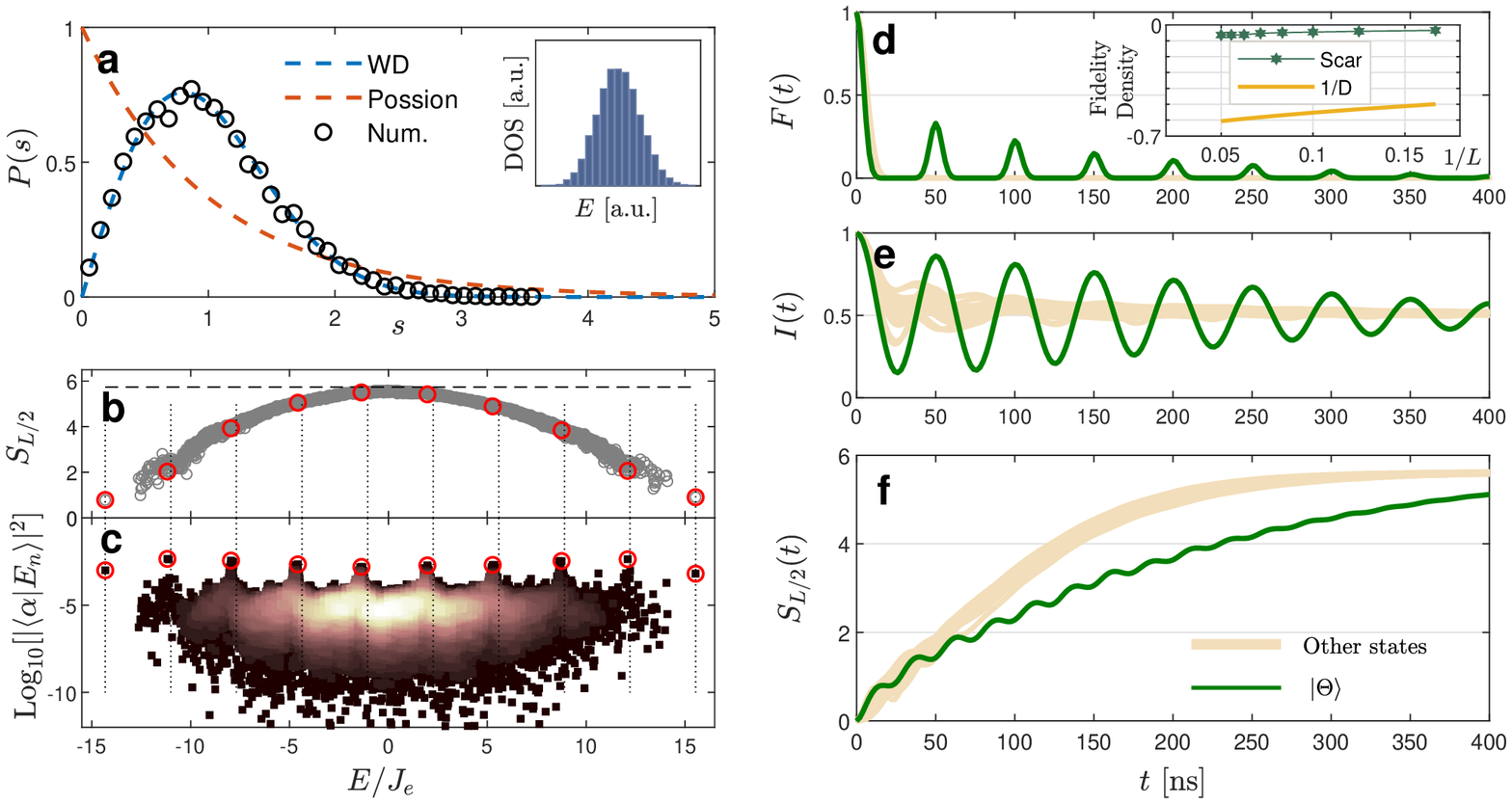,width=\linewidth}
\caption{ QMBS in a comb tensor configuration. The system parameters are $L=18$ and $N=9$, $J_\mathrm{a}/2\pi=1.5 J_\mathrm{e}/2\pi=-9$ MHz and $J_\mathrm{x}/2\pi\in[0.3,1.2]$ MHz,. (a) Level-spacing statistics, the black circles show the numerical data, while the dotted lines show the expected distributions for an integrable (Poisson) and a chaotic (Wigner-Dyson) model. The inset shows the density of states.
(b) Entanglement entropy of all eigenstates, with the horizontal dashed line
showing the approximate Page value $(L/2)\mathrm{ln}2-1/2$. (c) Squared overlap between the collective dimerized state $|\Theta\rangle$ and
the eigenstates $|E_n\rangle$. In (b) and (c), the red hexagrams mark the major eigenstates that contribute to the scarring
dynamics. (d-f) Time evolution of fidelity, imbalance, and entanglement entropy for the collective dimerized state (green)
and other random states (light orange). Inset of (d) shows the fidelity density for the $|\Theta\rangle$ state and the inverse Hilbert space dimension $1/D$ as a function of $1/L$.}
\label{sfig:scar_comb}
\end{figure*}

\begin{figure*} [ht!]
\centering
\epsfig{figure=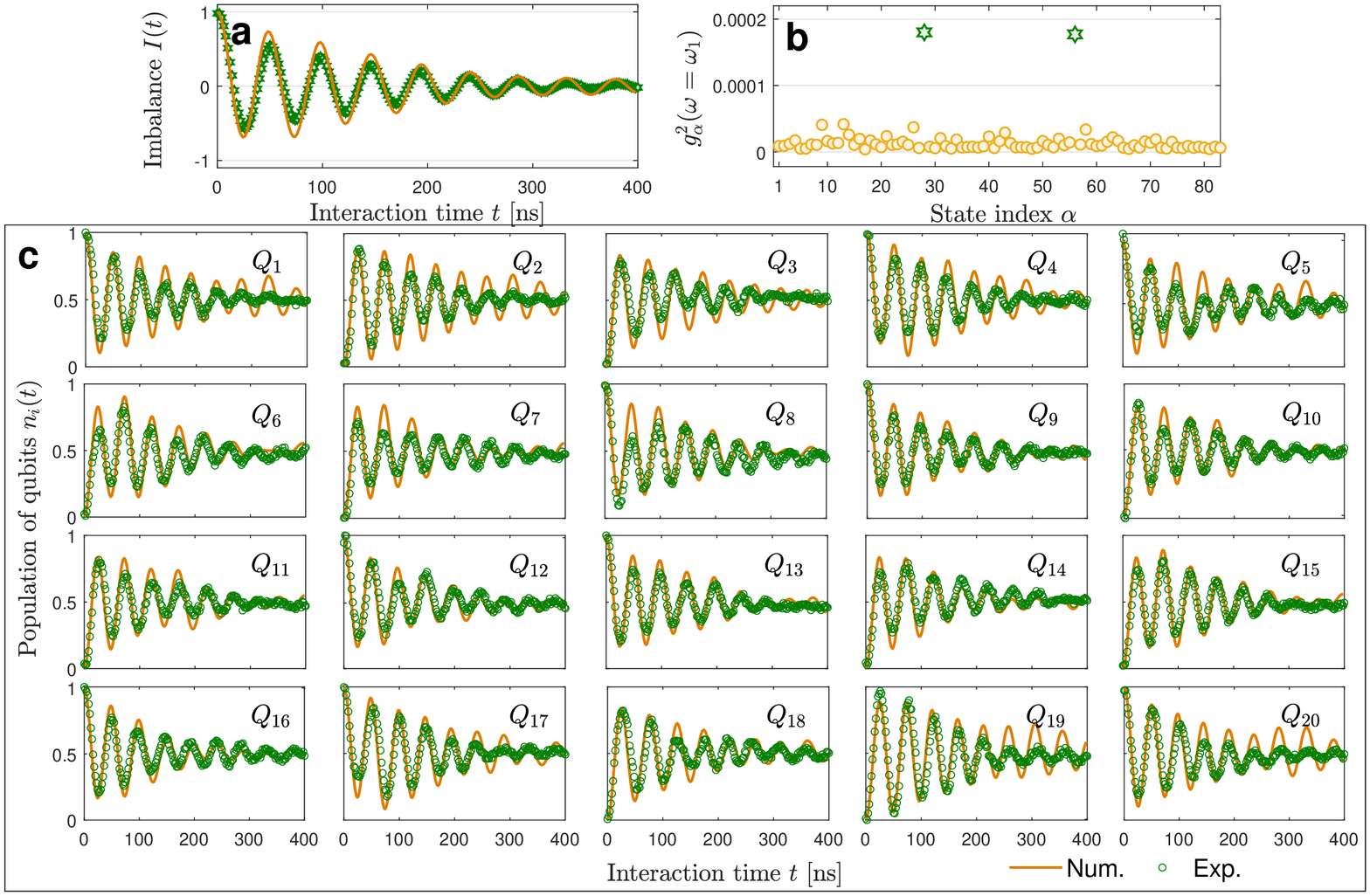,width=\linewidth}
\caption{Behaviors of imbalance of $|\Pi\rangle$. (a) Imbalance of $|\Pi\rangle$ as a function of the interaction time for $L=20$ from experimental measurement (green hexagrams) and numerical simulation (orange curve). (b) Squared Fourier amplitude at the revival frequency from randomly chosen initial product states. The green hexagrams show the $|\Pi\rangle$ and $|\Pi^\prime\rangle$ states. (c) Time evolution of qubit population for the $20$-qubit chain from experimental data (green circles) and simulation results (solid orange curves). The numerical parameter values are $J_\mathrm{a}/2\pi=-9$ MHz, $J_\mathrm{e}/2\pi=-6$ MHz, and $J_\mathrm{x}/2\pi$ is randomly chosen from the range $[0.3,1.2]$ MHz.}
\label{sfig:eachQ_ni}
\end{figure*}

\begin{figure*} [ht!]
\centering
\epsfig{figure=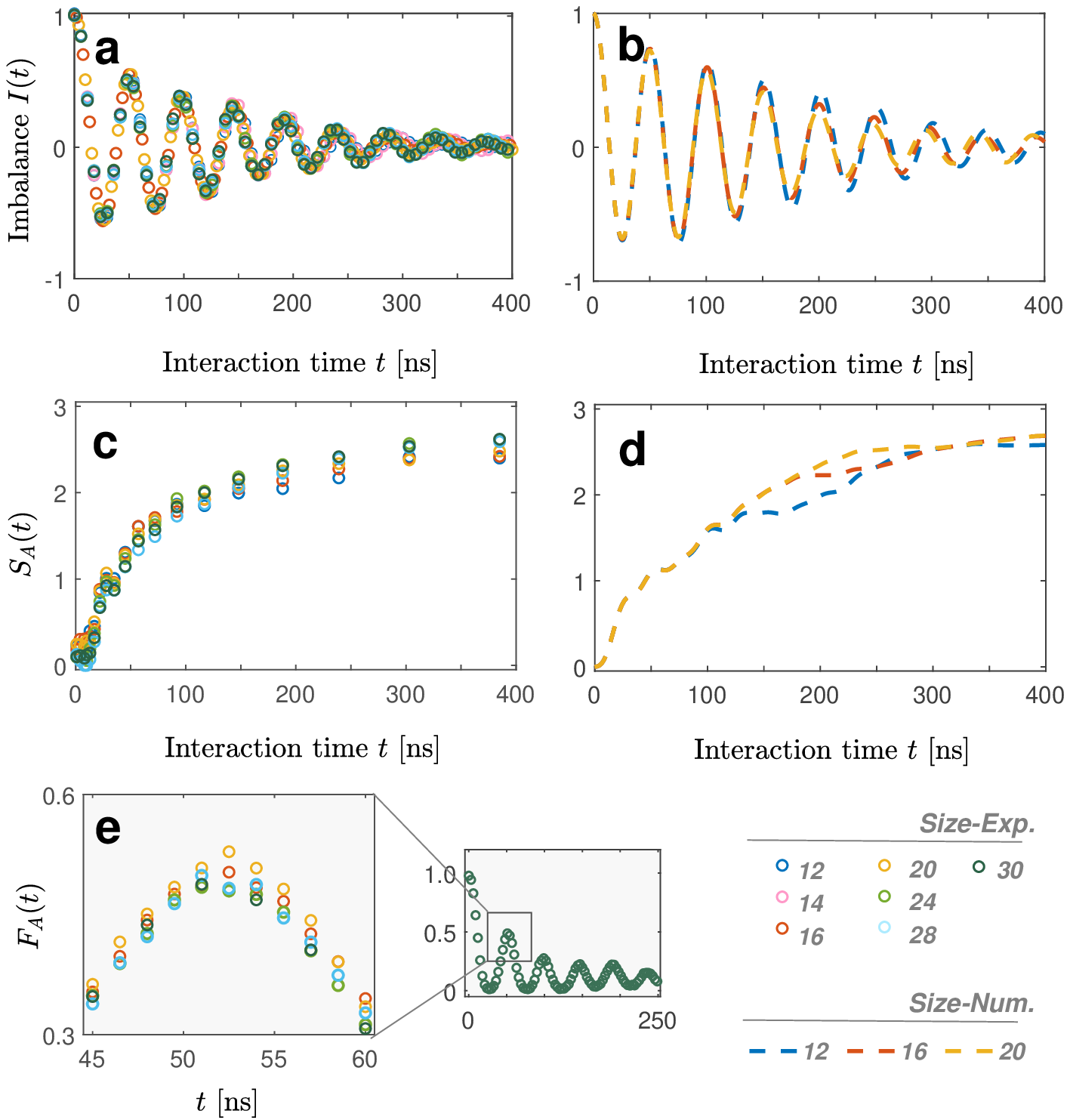,width=5.5in}
\caption{ Behaviors of imbalance of $|\Pi\rangle$ at different system sizes. (a,b) Experimentally measured and numerically calculated imbalance versus interaction time for system sizes ranging from $L=12$ to $30$. (c,d) Experimental and numerical four-qubit entanglement entropy versus interaction time for the same set of system sizes. For numerical simulations, the maximum size is $L=20$ due to computational constraints. (e) Four-qubit fidelity as a function of interaction time near the first revival.}
\label{sfig:scaling}
\end{figure*}

\begin{figure*} [ht!]
\centering
\epsfig{figure=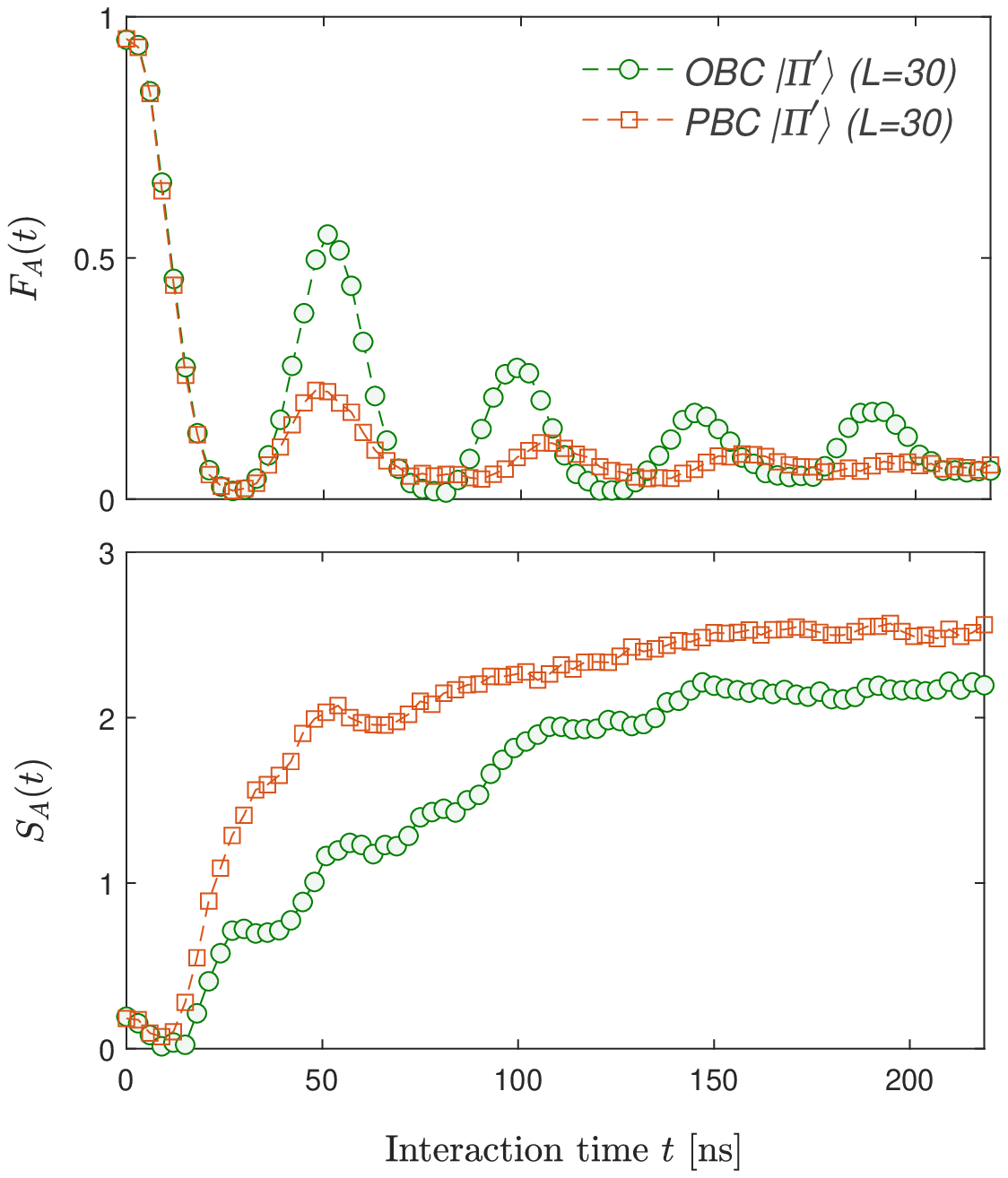,width=4in}
\caption{Comparison of four-qubit fidelity (upper) and entanglement entropy (lower) of scarred states for PBC and OBC. }
\label{sfig:pbc}
\end{figure*}


\begin{thebibliography}{56}%
\makeatletter
\providecommand \@ifxundefined [1]{%
 \@ifx{#1\undefined}
}%
\providecommand \@ifnum [1]{%
 \ifnum #1\expandafter \@firstoftwo
 \else \expandafter \@secondoftwo
 \fi
}%
\providecommand \@ifx [1]{%
 \ifx #1\expandafter \@firstoftwo
 \else \expandafter \@secondoftwo
 \fi
}%
\providecommand \natexlab [1]{#1}%
\providecommand \enquote  [1]{``#1''}%
\providecommand \bibnamefont  [1]{#1}%
\providecommand \bibfnamefont [1]{#1}%
\providecommand \citenamefont [1]{#1}%
\providecommand \href@noop [0]{\@secondoftwo}%
\providecommand \href [0]{\begingroup \@sanitize@url \@href}%
\providecommand \@href[1]{\@@startlink{#1}\@@href}%
\providecommand \@@href[1]{\endgroup#1\@@endlink}%
\providecommand \@sanitize@url [0]{\catcode `\\12\catcode `\$12\catcode
  `\&12\catcode `\#12\catcode `\^12\catcode `\_12\catcode `\%12\relax}%
\providecommand \@@startlink[1]{}%
\providecommand \@@endlink[0]{}%
\providecommand \url  [0]{\begingroup\@sanitize@url \@url }%
\providecommand \@url [1]{\endgroup\@href {#1}{\urlprefix }}%
\providecommand \urlprefix  [0]{URL }%
\providecommand \Eprint [0]{\href }%
\providecommand \doibase [0]{https://doi.org/}%
\providecommand \selectlanguage [0]{\@gobble}%
\providecommand \bibinfo  [0]{\@secondoftwo}%
\providecommand \bibfield  [0]{\@secondoftwo}%
\providecommand \translation [1]{[#1]}%
\providecommand \BibitemOpen [0]{}%
\providecommand \bibitemStop [0]{}%
\providecommand \bibitemNoStop [0]{.\EOS\space}%
\providecommand \EOS [0]{\spacefactor3000\relax}%
\providecommand \BibitemShut  [1]{\csname bibitem#1\endcsname}%
\let\auto@bib@innerbib\@empty
\bibitem [{\citenamefont {Ladd}\ \emph {et~al.}(2010)\citenamefont {Ladd},
  \citenamefont {Jelezko}, \citenamefont {Laflamme}, \citenamefont {Nakamura},
  \citenamefont {Monroe},\ and\ \citenamefont {O'Brien}}]{LJLN:2010}%
  \BibitemOpen
  \bibfield  {author} {\bibinfo {author} {\bibfnamefont {T.~D.}\ \bibnamefont
  {Ladd}}, \bibinfo {author} {\bibfnamefont {F.}~\bibnamefont {Jelezko}},
  \bibinfo {author} {\bibfnamefont {R.}~\bibnamefont {Laflamme}}, \bibinfo
  {author} {\bibfnamefont {Y.}~\bibnamefont {Nakamura}}, \bibinfo {author}
  {\bibfnamefont {C.}~\bibnamefont {Monroe}},\ and\ \bibinfo {author}
  {\bibfnamefont {J.~L.}\ \bibnamefont {O'Brien}},\ }\bibfield  {title}
  {\bibinfo {title} {Quantum computers},\ }\href
  {https://doi.org/10.1038/nature08812} {\bibfield  {journal} {\bibinfo
  {journal} {Nature}\ }\textbf {\bibinfo {volume} {464}},\ \bibinfo {pages}
  {45} (\bibinfo {year} {2010})}\BibitemShut {NoStop}%
\bibitem [{\citenamefont {Georgescu}\ \emph {et~al.}(2014)\citenamefont
  {Georgescu}, \citenamefont {Ashhab},\ and\ \citenamefont {Nori}}]{GAN:2014}%
  \BibitemOpen
  \bibfield  {author} {\bibinfo {author} {\bibfnamefont {I.~M.}\ \bibnamefont
  {Georgescu}}, \bibinfo {author} {\bibfnamefont {S.}~\bibnamefont {Ashhab}},\
  and\ \bibinfo {author} {\bibfnamefont {F.}~\bibnamefont {Nori}},\ }\bibfield
  {title} {\bibinfo {title} {Quantum simulation},\ }\href
  {https://doi.org/10.1103/RevModPhys.86.153} {\bibfield  {journal} {\bibinfo
  {journal} {Rev. Mod. Phys.}\ }\textbf {\bibinfo {volume} {86}},\ \bibinfo
  {pages} {153} (\bibinfo {year} {2014})}\BibitemShut {NoStop}%
\bibitem [{\citenamefont {Mi}\ \emph {et~al.}(2021)\citenamefont {Mi},
  \citenamefont {Roushan}, \citenamefont {Quintana}, \citenamefont {Mandra},
  \citenamefont {Marshall}, \citenamefont {Neill}, \citenamefont {Arute},
  \citenamefont {Arya}, \citenamefont {Atalaya}, \citenamefont {Babbush} \emph
  {et~al.}}]{MRQM:2021}%
  \BibitemOpen
  \bibfield  {author} {\bibinfo {author} {\bibfnamefont {X.}~\bibnamefont
  {Mi}}, \bibinfo {author} {\bibfnamefont {P.}~\bibnamefont {Roushan}},
  \bibinfo {author} {\bibfnamefont {C.}~\bibnamefont {Quintana}}, \bibinfo
  {author} {\bibfnamefont {S.}~\bibnamefont {Mandra}}, \bibinfo {author}
  {\bibfnamefont {J.}~\bibnamefont {Marshall}}, \bibinfo {author}
  {\bibfnamefont {C.}~\bibnamefont {Neill}}, \bibinfo {author} {\bibfnamefont
  {F.}~\bibnamefont {Arute}}, \bibinfo {author} {\bibfnamefont
  {K.}~\bibnamefont {Arya}}, \bibinfo {author} {\bibfnamefont {J.}~\bibnamefont
  {Atalaya}}, \bibinfo {author} {\bibfnamefont {R.}~\bibnamefont {Babbush}},
  \emph {et~al.},\ }\bibfield  {title} {\bibinfo {title} {Information
  scrambling in computationally complex quantum circuits},\ }\href
  {https://doi.org/10.1126/science.abg5029} {\bibfield  {journal} {\bibinfo
  {journal} {Science}\ }\textbf {\bibinfo {volume} {374}},\ \bibinfo {pages}
  {1479} (\bibinfo {year} {2021})}\BibitemShut {NoStop}%
\bibitem [{\citenamefont {Arute}\ \emph {et~al.}(2019)\citenamefont {Arute},
  \citenamefont {Arya}, \citenamefont {Babbush}, \citenamefont {Bacon},
  \citenamefont {Bardin}, \citenamefont {Barends}, \citenamefont {Biswas},
  \citenamefont {Boixo}, \citenamefont {Brandao}, \citenamefont {Buell} \emph
  {et~al.}}]{AABR:2019}%
  \BibitemOpen
  \bibfield  {author} {\bibinfo {author} {\bibfnamefont {F.}~\bibnamefont
  {Arute}}, \bibinfo {author} {\bibfnamefont {K.}~\bibnamefont {Arya}},
  \bibinfo {author} {\bibfnamefont {R.}~\bibnamefont {Babbush}}, \bibinfo
  {author} {\bibfnamefont {D.}~\bibnamefont {Bacon}}, \bibinfo {author}
  {\bibfnamefont {J.~C.}\ \bibnamefont {Bardin}}, \bibinfo {author}
  {\bibfnamefont {R.}~\bibnamefont {Barends}}, \bibinfo {author} {\bibfnamefont
  {R.}~\bibnamefont {Biswas}}, \bibinfo {author} {\bibfnamefont
  {S.}~\bibnamefont {Boixo}}, \bibinfo {author} {\bibfnamefont {F.~G.}\
  \bibnamefont {Brandao}}, \bibinfo {author} {\bibfnamefont {D.~A.}\
  \bibnamefont {Buell}}, \emph {et~al.},\ }\bibfield  {title} {\bibinfo {title}
  {Quantum supremacy using a programmable superconducting processor},\ }\href
  {https://doi.org/10.1038/s41586-019-1666-5} {\bibfield  {journal} {\bibinfo
  {journal} {Nature}\ }\textbf {\bibinfo {volume} {574}},\ \bibinfo {pages}
  {505} (\bibinfo {year} {2019})}\BibitemShut {NoStop}%
\bibitem [{\citenamefont {Swingle}\ \emph {et~al.}(2016)\citenamefont
  {Swingle}, \citenamefont {Bentsen}, \citenamefont {Schleier-Smith},\ and\
  \citenamefont {Hayden}}]{SBSH:2016}%
  \BibitemOpen
  \bibfield  {author} {\bibinfo {author} {\bibfnamefont {B.}~\bibnamefont
  {Swingle}}, \bibinfo {author} {\bibfnamefont {G.}~\bibnamefont {Bentsen}},
  \bibinfo {author} {\bibfnamefont {M.}~\bibnamefont {Schleier-Smith}},\ and\
  \bibinfo {author} {\bibfnamefont {P.}~\bibnamefont {Hayden}},\ }\bibfield
  {title} {\bibinfo {title} {Measuring the scrambling of quantum information},\
  }\href {https://doi.org/10.1103/PhysRevA.94.040302} {\bibfield  {journal}
  {\bibinfo  {journal} {Phys. Rev. A}\ }\textbf {\bibinfo {volume} {94}},\
  \bibinfo {pages} {040302} (\bibinfo {year} {2016})}\BibitemShut {NoStop}%
\bibitem [{\citenamefont {Xu}\ \emph {et~al.}(2018)\citenamefont {Xu},
  \citenamefont {Chen}, \citenamefont {Zeng}, \citenamefont {Zhang},
  \citenamefont {Song}, \citenamefont {Liu}, \citenamefont {Guo}, \citenamefont
  {Zhang}, \citenamefont {Xu}, \citenamefont {Deng}, \citenamefont {Huang},
  \citenamefont {Wang}, \citenamefont {Zhu}, \citenamefont {Zheng},\ and\
  \citenamefont {Fan}}]{XCZZ:2018}%
  \BibitemOpen
  \bibfield  {author} {\bibinfo {author} {\bibfnamefont {K.}~\bibnamefont
  {Xu}}, \bibinfo {author} {\bibfnamefont {J.-J.}\ \bibnamefont {Chen}},
  \bibinfo {author} {\bibfnamefont {Y.}~\bibnamefont {Zeng}}, \bibinfo {author}
  {\bibfnamefont {Y.-R.}\ \bibnamefont {Zhang}}, \bibinfo {author}
  {\bibfnamefont {C.}~\bibnamefont {Song}}, \bibinfo {author} {\bibfnamefont
  {W.}~\bibnamefont {Liu}}, \bibinfo {author} {\bibfnamefont {Q.}~\bibnamefont
  {Guo}}, \bibinfo {author} {\bibfnamefont {P.}~\bibnamefont {Zhang}}, \bibinfo
  {author} {\bibfnamefont {D.}~\bibnamefont {Xu}}, \bibinfo {author}
  {\bibfnamefont {H.}~\bibnamefont {Deng}}, \bibinfo {author} {\bibfnamefont
  {K.}~\bibnamefont {Huang}}, \bibinfo {author} {\bibfnamefont
  {H.}~\bibnamefont {Wang}}, \bibinfo {author} {\bibfnamefont {X.}~\bibnamefont
  {Zhu}}, \bibinfo {author} {\bibfnamefont {D.}~\bibnamefont {Zheng}},\ and\
  \bibinfo {author} {\bibfnamefont {H.}~\bibnamefont {Fan}},\ }\bibfield
  {title} {\bibinfo {title} {Emulating many-body localization with a
  superconducting quantum processor},\ }\href
  {https://doi.org/10.1103/PhysRevLett.120.050507} {\bibfield  {journal}
  {\bibinfo  {journal} {Phys. Rev. Lett.}\ }\textbf {\bibinfo {volume} {120}},\
  \bibinfo {pages} {050507} (\bibinfo {year} {2018})}\BibitemShut {NoStop}%
\bibitem [{\citenamefont {Landsman}\ \emph {et~al.}(2019)\citenamefont
  {Landsman}, \citenamefont {Figgatt}, \citenamefont {Schuster}, \citenamefont
  {Linke}, \citenamefont {Yoshida}, \citenamefont {Yao},\ and\ \citenamefont
  {Monroe}}]{LFSL:2019}%
  \BibitemOpen
  \bibfield  {author} {\bibinfo {author} {\bibfnamefont {K.~A.}\ \bibnamefont
  {Landsman}}, \bibinfo {author} {\bibfnamefont {C.}~\bibnamefont {Figgatt}},
  \bibinfo {author} {\bibfnamefont {T.}~\bibnamefont {Schuster}}, \bibinfo
  {author} {\bibfnamefont {N.~M.}\ \bibnamefont {Linke}}, \bibinfo {author}
  {\bibfnamefont {B.}~\bibnamefont {Yoshida}}, \bibinfo {author} {\bibfnamefont
  {N.~Y.}\ \bibnamefont {Yao}},\ and\ \bibinfo {author} {\bibfnamefont
  {C.}~\bibnamefont {Monroe}},\ }\bibfield  {title} {\bibinfo {title} {Verified
  quantum information scrambling},\ }\href
  {https://doi.org/10.1038/s41586-019-0952-6} {\bibfield  {journal} {\bibinfo
  {journal} {Nature}\ }\textbf {\bibinfo {volume} {567}},\ \bibinfo {pages}
  {61} (\bibinfo {year} {2019})}\BibitemShut {NoStop}%
\bibitem [{\citenamefont {Morong}\ \emph {et~al.}(2021)\citenamefont {Morong},
  \citenamefont {Liu}, \citenamefont {Becker}, \citenamefont {Collins},
  \citenamefont {Feng}, \citenamefont {Kyprianidis}, \citenamefont {Pagano},
  \citenamefont {You}, \citenamefont {Gorshkov},\ and\ \citenamefont
  {Monroe}}]{MLBC:2021}%
  \BibitemOpen
  \bibfield  {author} {\bibinfo {author} {\bibfnamefont {W.}~\bibnamefont
  {Morong}}, \bibinfo {author} {\bibfnamefont {F.}~\bibnamefont {Liu}},
  \bibinfo {author} {\bibfnamefont {P.}~\bibnamefont {Becker}}, \bibinfo
  {author} {\bibfnamefont {K.~S.}\ \bibnamefont {Collins}}, \bibinfo {author}
  {\bibfnamefont {L.}~\bibnamefont {Feng}}, \bibinfo {author} {\bibfnamefont
  {A.}~\bibnamefont {Kyprianidis}}, \bibinfo {author} {\bibfnamefont
  {G.}~\bibnamefont {Pagano}}, \bibinfo {author} {\bibfnamefont
  {T.}~\bibnamefont {You}}, \bibinfo {author} {\bibfnamefont {A.~V.}\
  \bibnamefont {Gorshkov}},\ and\ \bibinfo {author} {\bibfnamefont
  {C.}~\bibnamefont {Monroe}},\ }\bibfield  {title} {\bibinfo {title}
  {Observation of {Stark} many-body localization without disorder},\ }\href
  {https://doi.org/10.1038/s41586-021-03988-0} {\bibfield  {journal} {\bibinfo
  {journal} {Nature}\ }\textbf {\bibinfo {volume} {599}},\ \bibinfo {pages}
  {393} (\bibinfo {year} {2021})}\BibitemShut {NoStop}%
\bibitem [{\citenamefont {Deutsch}(1991)}]{Deutsch:1991}%
  \BibitemOpen
  \bibfield  {author} {\bibinfo {author} {\bibfnamefont {J.~M.}\ \bibnamefont
  {Deutsch}},\ }\bibfield  {title} {\bibinfo {title} {Quantum statistical
  mechanics in a closed system},\ }\href
  {https://doi.org/10.1103/PhysRevA.43.2046} {\bibfield  {journal} {\bibinfo
  {journal} {Phys. Rev. A}\ }\textbf {\bibinfo {volume} {43}},\ \bibinfo
  {pages} {2046} (\bibinfo {year} {1991})}\BibitemShut {NoStop}%
\bibitem [{\citenamefont {Srednicki}(1994)}]{Srednicki:1994}%
  \BibitemOpen
  \bibfield  {author} {\bibinfo {author} {\bibfnamefont {M.}~\bibnamefont
  {Srednicki}},\ }\bibfield  {title} {\bibinfo {title} {Chaos and quantum
  thermalization},\ }\href {https://doi.org/10.1103/PhysRevE.50.888} {\bibfield
   {journal} {\bibinfo  {journal} {Phys. Rev. E}\ }\textbf {\bibinfo {volume}
  {50}},\ \bibinfo {pages} {888} (\bibinfo {year} {1994})}\BibitemShut
  {NoStop}%
\bibitem [{\citenamefont {Rigol}\ \emph {et~al.}(2008)\citenamefont {Rigol},
  \citenamefont {Dunjko},\ and\ \citenamefont {Olshanii}}]{RDO:2008}%
  \BibitemOpen
  \bibfield  {author} {\bibinfo {author} {\bibfnamefont {M.}~\bibnamefont
  {Rigol}}, \bibinfo {author} {\bibfnamefont {V.}~\bibnamefont {Dunjko}},\ and\
  \bibinfo {author} {\bibfnamefont {M.}~\bibnamefont {Olshanii}},\ }\bibfield
  {title} {\bibinfo {title} {Thermalization and its mechanism for generic
  isolated quantum systems},\ }\href {https://doi.org/10.1038/nature06838}
  {\bibfield  {journal} {\bibinfo  {journal} {Nature}\ }\textbf {\bibinfo
  {volume} {452}},\ \bibinfo {pages} {854} (\bibinfo {year}
  {2008})}\BibitemShut {NoStop}%
\bibitem [{\citenamefont {Kaufman}\ \emph {et~al.}(2016)\citenamefont
  {Kaufman}, \citenamefont {Tai}, \citenamefont {Lukin}, \citenamefont
  {Rispoli}, \citenamefont {Schittko}, \citenamefont {Preiss},\ and\
  \citenamefont {Greiner}}]{KTLR:2016}%
  \BibitemOpen
  \bibfield  {author} {\bibinfo {author} {\bibfnamefont {A.~M.}\ \bibnamefont
  {Kaufman}}, \bibinfo {author} {\bibfnamefont {M.~E.}\ \bibnamefont {Tai}},
  \bibinfo {author} {\bibfnamefont {A.}~\bibnamefont {Lukin}}, \bibinfo
  {author} {\bibfnamefont {M.}~\bibnamefont {Rispoli}}, \bibinfo {author}
  {\bibfnamefont {R.}~\bibnamefont {Schittko}}, \bibinfo {author}
  {\bibfnamefont {P.~M.}\ \bibnamefont {Preiss}},\ and\ \bibinfo {author}
  {\bibfnamefont {M.}~\bibnamefont {Greiner}},\ }\bibfield  {title} {\bibinfo
  {title} {Quantum thermalization through entanglement in an isolated many-body
  system},\ }\href {https://doi.org/10.1126/science.aaf6725} {\bibfield
  {journal} {\bibinfo  {journal} {Science}\ }\textbf {\bibinfo {volume}
  {353}},\ \bibinfo {pages} {794} (\bibinfo {year} {2016})}\BibitemShut
  {NoStop}%
\bibitem [{\citenamefont {Bluvstein}\ \emph {et~al.}(2021)\citenamefont
  {Bluvstein}, \citenamefont {Omran}, \citenamefont {Levine}, \citenamefont
  {Keesling}, \citenamefont {Semeghini}, \citenamefont {Ebadi}, \citenamefont
  {Wang}, \citenamefont {Michailidis}, \citenamefont {Maskara}, \citenamefont
  {Ho}, \citenamefont {Choi}, \citenamefont {Serbyn}, \citenamefont {Greiner},
  \citenamefont {Vuletić},\ and\ \citenamefont {Lukin}}]{BOLK:2021}%
  \BibitemOpen
  \bibfield  {author} {\bibinfo {author} {\bibfnamefont {D.}~\bibnamefont
  {Bluvstein}}, \bibinfo {author} {\bibfnamefont {A.}~\bibnamefont {Omran}},
  \bibinfo {author} {\bibfnamefont {H.}~\bibnamefont {Levine}}, \bibinfo
  {author} {\bibfnamefont {A.}~\bibnamefont {Keesling}}, \bibinfo {author}
  {\bibfnamefont {G.}~\bibnamefont {Semeghini}}, \bibinfo {author}
  {\bibfnamefont {S.}~\bibnamefont {Ebadi}}, \bibinfo {author} {\bibfnamefont
  {T.~T.}\ \bibnamefont {Wang}}, \bibinfo {author} {\bibfnamefont {A.~A.}\
  \bibnamefont {Michailidis}}, \bibinfo {author} {\bibfnamefont
  {N.}~\bibnamefont {Maskara}}, \bibinfo {author} {\bibfnamefont {W.~W.}\
  \bibnamefont {Ho}}, \bibinfo {author} {\bibfnamefont {S.}~\bibnamefont
  {Choi}}, \bibinfo {author} {\bibfnamefont {M.}~\bibnamefont {Serbyn}},
  \bibinfo {author} {\bibfnamefont {M.}~\bibnamefont {Greiner}}, \bibinfo
  {author} {\bibfnamefont {V.}~\bibnamefont {Vuletić}},\ and\ \bibinfo
  {author} {\bibfnamefont {M.~D.}\ \bibnamefont {Lukin}},\ }\bibfield  {title}
  {\bibinfo {title} {Controlling quantum many-body dynamics in driven {Rydberg}
  atom arrays},\ }\href {https://doi.org/10.1126/science.abg2530} {\bibfield
  {journal} {\bibinfo  {journal} {Science}\ }\textbf {\bibinfo {volume}
  {371}},\ \bibinfo {pages} {1355} (\bibinfo {year} {2021})}\BibitemShut
  {NoStop}%
\bibitem [{\citenamefont {Serbyn}\ \emph {et~al.}(2021)\citenamefont {Serbyn},
  \citenamefont {Abanin},\ and\ \citenamefont {Papi{\'{c}}}}]{SAP:2021}%
  \BibitemOpen
  \bibfield  {author} {\bibinfo {author} {\bibfnamefont {M.}~\bibnamefont
  {Serbyn}}, \bibinfo {author} {\bibfnamefont {D.~A.}\ \bibnamefont {Abanin}},\
  and\ \bibinfo {author} {\bibfnamefont {Z.}~\bibnamefont {Papi{\'{c}}}},\
  }\bibfield  {title} {\bibinfo {title} {Quantum many-body scars and weak
  breaking of ergodicity},\ }\href {https://doi.org/10.1038/s41567-021-01230-2}
  {\bibfield  {journal} {\bibinfo  {journal} {Nature Physics}\ }\textbf
  {\bibinfo {volume} {17}},\ \bibinfo {pages} {675} (\bibinfo {year}
  {2021})}\BibitemShut {NoStop}%
\bibitem [{\citenamefont {Moudgalya}\ \emph {et~al.}(2021)\citenamefont
  {Moudgalya}, \citenamefont {Bernevig},\ and\ \citenamefont
  {Regnault}}]{MoudgalyaReview}%
  \BibitemOpen
  \bibfield  {author} {\bibinfo {author} {\bibfnamefont {S.}~\bibnamefont
  {Moudgalya}}, \bibinfo {author} {\bibfnamefont {B.~A.}\ \bibnamefont
  {Bernevig}},\ and\ \bibinfo {author} {\bibfnamefont {N.}~\bibnamefont
  {Regnault}},\ }\href@noop {} {\bibinfo {title} {Quantum many-body scars and
  {Hilbert} space fragmentation: A review of exact results}} (\bibinfo {year}
  {2021}),\ \Eprint {https://arxiv.org/abs/2109.00548} {arXiv:2109.00548
  [cond-mat.str-el]} \BibitemShut {NoStop}%
\bibitem [{\citenamefont {Nandkishore}\ and\ \citenamefont
  {Huse}(2015)}]{NH:2015}%
  \BibitemOpen
  \bibfield  {author} {\bibinfo {author} {\bibfnamefont {R.}~\bibnamefont
  {Nandkishore}}\ and\ \bibinfo {author} {\bibfnamefont {D.~A.}\ \bibnamefont
  {Huse}},\ }\bibfield  {title} {\bibinfo {title} {Many-body localization and
  thermalization in quantum statistical mechanics},\ }\href
  {https://doi.org/10.1146/annurev-conmatphys-031214-014726} {\bibfield
  {journal} {\bibinfo  {journal} {Annual Review of Condensed Matter Physics}\
  }\textbf {\bibinfo {volume} {6}},\ \bibinfo {pages} {15} (\bibinfo {year}
  {2015})}\BibitemShut {NoStop}%
\bibitem [{\citenamefont {Abanin}\ \emph {et~al.}(2019)\citenamefont {Abanin},
  \citenamefont {Altman}, \citenamefont {Bloch},\ and\ \citenamefont
  {Serbyn}}]{AABS:2019}%
  \BibitemOpen
  \bibfield  {author} {\bibinfo {author} {\bibfnamefont {D.~A.}\ \bibnamefont
  {Abanin}}, \bibinfo {author} {\bibfnamefont {E.}~\bibnamefont {Altman}},
  \bibinfo {author} {\bibfnamefont {I.}~\bibnamefont {Bloch}},\ and\ \bibinfo
  {author} {\bibfnamefont {M.}~\bibnamefont {Serbyn}},\ }\bibfield  {title}
  {\bibinfo {title} {Colloquium: Many-body localization, thermalization, and
  entanglement},\ }\href {https://doi.org/10.1103/RevModPhys.91.021001}
  {\bibfield  {journal} {\bibinfo  {journal} {Rev. Mod. Phys.}\ }\textbf
  {\bibinfo {volume} {91}},\ \bibinfo {pages} {021001} (\bibinfo {year}
  {2019})}\BibitemShut {NoStop}%
\bibitem [{\citenamefont {Guo}\ \emph {et~al.}(2021{\natexlab{a}})\citenamefont
  {Guo}, \citenamefont {Cheng}, \citenamefont {Li}, \citenamefont {Xu},
  \citenamefont {Zhang}, \citenamefont {Wang}, \citenamefont {Song},
  \citenamefont {Liu}, \citenamefont {Ren}, \citenamefont {Dong}, \citenamefont
  {Mondaini},\ and\ \citenamefont {Wang}}]{GCLX:2021}%
  \BibitemOpen
  \bibfield  {author} {\bibinfo {author} {\bibfnamefont {Q.}~\bibnamefont
  {Guo}}, \bibinfo {author} {\bibfnamefont {C.}~\bibnamefont {Cheng}}, \bibinfo
  {author} {\bibfnamefont {H.}~\bibnamefont {Li}}, \bibinfo {author}
  {\bibfnamefont {S.}~\bibnamefont {Xu}}, \bibinfo {author} {\bibfnamefont
  {P.}~\bibnamefont {Zhang}}, \bibinfo {author} {\bibfnamefont
  {Z.}~\bibnamefont {Wang}}, \bibinfo {author} {\bibfnamefont {C.}~\bibnamefont
  {Song}}, \bibinfo {author} {\bibfnamefont {W.}~\bibnamefont {Liu}}, \bibinfo
  {author} {\bibfnamefont {W.}~\bibnamefont {Ren}}, \bibinfo {author}
  {\bibfnamefont {H.}~\bibnamefont {Dong}}, \bibinfo {author} {\bibfnamefont
  {R.}~\bibnamefont {Mondaini}},\ and\ \bibinfo {author} {\bibfnamefont
  {H.}~\bibnamefont {Wang}},\ }\bibfield  {title} {\bibinfo {title} {Stark
  many-body localization on a superconducting quantum processor},\ }\href
  {https://doi.org/10.1103/PhysRevLett.127.240502} {\bibfield  {journal}
  {\bibinfo  {journal} {Phys. Rev. Lett.}\ }\textbf {\bibinfo {volume} {127}},\
  \bibinfo {pages} {240502} (\bibinfo {year} {2021}{\natexlab{a}})}\BibitemShut
  {NoStop}%
\bibitem [{\citenamefont {Guo}\ \emph {et~al.}(2021{\natexlab{b}})\citenamefont
  {Guo}, \citenamefont {Cheng}, \citenamefont {Sun}, \citenamefont {Song},
  \citenamefont {Li}, \citenamefont {Wang}, \citenamefont {Ren}, \citenamefont
  {Dong}, \citenamefont {Zheng}, \citenamefont {Zhang} \emph
  {et~al.}}]{GCSS:2021}%
  \BibitemOpen
  \bibfield  {author} {\bibinfo {author} {\bibfnamefont {Q.}~\bibnamefont
  {Guo}}, \bibinfo {author} {\bibfnamefont {C.}~\bibnamefont {Cheng}}, \bibinfo
  {author} {\bibfnamefont {Z.-H.}\ \bibnamefont {Sun}}, \bibinfo {author}
  {\bibfnamefont {Z.}~\bibnamefont {Song}}, \bibinfo {author} {\bibfnamefont
  {H.}~\bibnamefont {Li}}, \bibinfo {author} {\bibfnamefont {Z.}~\bibnamefont
  {Wang}}, \bibinfo {author} {\bibfnamefont {W.}~\bibnamefont {Ren}}, \bibinfo
  {author} {\bibfnamefont {H.}~\bibnamefont {Dong}}, \bibinfo {author}
  {\bibfnamefont {D.}~\bibnamefont {Zheng}}, \bibinfo {author} {\bibfnamefont
  {Y.-R.}\ \bibnamefont {Zhang}}, \emph {et~al.},\ }\bibfield  {title}
  {\bibinfo {title} {Observation of energy-resolved many-body localization},\
  }\href@noop {} {\bibfield  {journal} {\bibinfo  {journal} {Nat. Phys.}\
  }\textbf {\bibinfo {volume} {17}},\ \bibinfo {pages} {234} (\bibinfo {year}
  {2021}{\natexlab{b}})}\BibitemShut {NoStop}%
\bibitem [{\citenamefont {Omran}\ \emph {et~al.}(2019)\citenamefont {Omran},
  \citenamefont {Levine}, \citenamefont {Keesling}, \citenamefont {Semeghini},
  \citenamefont {Wang}, \citenamefont {Ebadi}, \citenamefont {Bernien},
  \citenamefont {Zibrov}, \citenamefont {Pichler}, \citenamefont {Choi},
  \citenamefont {Cui}, \citenamefont {Rossignolo}, \citenamefont {Rembold},
  \citenamefont {Montangero}, \citenamefont {Calarco}, \citenamefont {Endres},
  \citenamefont {Greiner}, \citenamefont {Vuletić},\ and\ \citenamefont
  {Lukin}}]{OLKS:2019}%
  \BibitemOpen
  \bibfield  {author} {\bibinfo {author} {\bibfnamefont {A.}~\bibnamefont
  {Omran}}, \bibinfo {author} {\bibfnamefont {H.}~\bibnamefont {Levine}},
  \bibinfo {author} {\bibfnamefont {A.}~\bibnamefont {Keesling}}, \bibinfo
  {author} {\bibfnamefont {G.}~\bibnamefont {Semeghini}}, \bibinfo {author}
  {\bibfnamefont {T.~T.}\ \bibnamefont {Wang}}, \bibinfo {author}
  {\bibfnamefont {S.}~\bibnamefont {Ebadi}}, \bibinfo {author} {\bibfnamefont
  {H.}~\bibnamefont {Bernien}}, \bibinfo {author} {\bibfnamefont {A.~S.}\
  \bibnamefont {Zibrov}}, \bibinfo {author} {\bibfnamefont {H.}~\bibnamefont
  {Pichler}}, \bibinfo {author} {\bibfnamefont {S.}~\bibnamefont {Choi}},
  \bibinfo {author} {\bibfnamefont {J.}~\bibnamefont {Cui}}, \bibinfo {author}
  {\bibfnamefont {M.}~\bibnamefont {Rossignolo}}, \bibinfo {author}
  {\bibfnamefont {P.}~\bibnamefont {Rembold}}, \bibinfo {author} {\bibfnamefont
  {S.}~\bibnamefont {Montangero}}, \bibinfo {author} {\bibfnamefont
  {T.}~\bibnamefont {Calarco}}, \bibinfo {author} {\bibfnamefont
  {M.}~\bibnamefont {Endres}}, \bibinfo {author} {\bibfnamefont
  {M.}~\bibnamefont {Greiner}}, \bibinfo {author} {\bibfnamefont
  {V.}~\bibnamefont {Vuletić}},\ and\ \bibinfo {author} {\bibfnamefont
  {M.~D.}\ \bibnamefont {Lukin}},\ }\bibfield  {title} {\bibinfo {title}
  {Generation and manipulation of {Schr\"odinger} cat states in {Rydberg} atom
  arrays},\ }\href {https://doi.org/10.1126/science.aax9743} {\bibfield
  {journal} {\bibinfo  {journal} {Science}\ }\textbf {\bibinfo {volume}
  {365}},\ \bibinfo {pages} {570} (\bibinfo {year} {2019})}\BibitemShut
  {NoStop}%
\bibitem [{\citenamefont {Dooley}(2021)}]{Dooley:2021}%
  \BibitemOpen
  \bibfield  {author} {\bibinfo {author} {\bibfnamefont {S.}~\bibnamefont
  {Dooley}},\ }\bibfield  {title} {\bibinfo {title} {Robust quantum sensing in
  strongly interacting systems with many-body scars},\ }\href
  {https://doi.org/10.1103/PRXQuantum.2.020330} {\bibfield  {journal} {\bibinfo
   {journal} {PRX Quantum}\ }\textbf {\bibinfo {volume} {2}},\ \bibinfo {pages}
  {020330} (\bibinfo {year} {2021})}\BibitemShut {NoStop}%
\bibitem [{\citenamefont {Shiraishi}\ and\ \citenamefont
  {Mori}(2017)}]{SM:2017}%
  \BibitemOpen
  \bibfield  {author} {\bibinfo {author} {\bibfnamefont {N.}~\bibnamefont
  {Shiraishi}}\ and\ \bibinfo {author} {\bibfnamefont {T.}~\bibnamefont
  {Mori}},\ }\bibfield  {title} {\bibinfo {title} {Systematic construction of
  counterexamples to the eigenstate thermalization hypothesis},\ }\href
  {https://doi.org/10.1103/PhysRevLett.119.030601} {\bibfield  {journal}
  {\bibinfo  {journal} {Phys. Rev. Lett.}\ }\textbf {\bibinfo {volume} {119}},\
  \bibinfo {pages} {030601} (\bibinfo {year} {2017})}\BibitemShut {NoStop}%
\bibitem [{\citenamefont {Moudgalya}\ \emph {et~al.}(2018)\citenamefont
  {Moudgalya}, \citenamefont {Regnault},\ and\ \citenamefont
  {Bernevig}}]{MRB:2018}%
  \BibitemOpen
  \bibfield  {author} {\bibinfo {author} {\bibfnamefont {S.}~\bibnamefont
  {Moudgalya}}, \bibinfo {author} {\bibfnamefont {N.}~\bibnamefont
  {Regnault}},\ and\ \bibinfo {author} {\bibfnamefont {B.~A.}\ \bibnamefont
  {Bernevig}},\ }\bibfield  {title} {\bibinfo {title} {Entanglement of exact
  excited states of {Affleck-Kennedy-Lieb-Tasaki} models: Exact results,
  many-body scars, and violation of the strong eigenstate thermalization
  hypothesis},\ }\href {https://doi.org/10.1103/PhysRevB.98.235156} {\bibfield
  {journal} {\bibinfo  {journal} {Phys. Rev. B}\ }\textbf {\bibinfo {volume}
  {98}},\ \bibinfo {pages} {235156} (\bibinfo {year} {2018})}\BibitemShut
  {NoStop}%
\bibitem [{\citenamefont {Ok}\ \emph {et~al.}(2019)\citenamefont {Ok},
  \citenamefont {Choo}, \citenamefont {Mudry}, \citenamefont {Castelnovo},
  \citenamefont {Chamon},\ and\ \citenamefont {Neupert}}]{OCMC:2019}%
  \BibitemOpen
  \bibfield  {author} {\bibinfo {author} {\bibfnamefont {S.}~\bibnamefont
  {Ok}}, \bibinfo {author} {\bibfnamefont {K.}~\bibnamefont {Choo}}, \bibinfo
  {author} {\bibfnamefont {C.}~\bibnamefont {Mudry}}, \bibinfo {author}
  {\bibfnamefont {C.}~\bibnamefont {Castelnovo}}, \bibinfo {author}
  {\bibfnamefont {C.}~\bibnamefont {Chamon}},\ and\ \bibinfo {author}
  {\bibfnamefont {T.}~\bibnamefont {Neupert}},\ }\bibfield  {title} {\bibinfo
  {title} {Topological many-body scar states in dimensions one, two, and
  three},\ }\href {https://doi.org/10.1103/PhysRevResearch.1.033144} {\bibfield
   {journal} {\bibinfo  {journal} {Phys. Rev. Research}\ }\textbf {\bibinfo
  {volume} {1}},\ \bibinfo {pages} {033144} (\bibinfo {year}
  {2019})}\BibitemShut {NoStop}%
\bibitem [{\citenamefont {Schecter}\ and\ \citenamefont
  {Iadecola}(2019)}]{SI:2019}%
  \BibitemOpen
  \bibfield  {author} {\bibinfo {author} {\bibfnamefont {M.}~\bibnamefont
  {Schecter}}\ and\ \bibinfo {author} {\bibfnamefont {T.}~\bibnamefont
  {Iadecola}},\ }\bibfield  {title} {\bibinfo {title} {Weak ergodicity breaking
  and quantum many-body scars in spin-1 {$XY$} magnets},\ }\href
  {https://doi.org/10.1103/PhysRevLett.123.147201} {\bibfield  {journal}
  {\bibinfo  {journal} {Phys. Rev. Lett.}\ }\textbf {\bibinfo {volume} {123}},\
  \bibinfo {pages} {147201} (\bibinfo {year} {2019})}\BibitemShut {NoStop}%
\bibitem [{\citenamefont {Shibata}\ \emph {et~al.}(2020)\citenamefont
  {Shibata}, \citenamefont {Yoshioka},\ and\ \citenamefont
  {Katsura}}]{SYK:2020}%
  \BibitemOpen
  \bibfield  {author} {\bibinfo {author} {\bibfnamefont {N.}~\bibnamefont
  {Shibata}}, \bibinfo {author} {\bibfnamefont {N.}~\bibnamefont {Yoshioka}},\
  and\ \bibinfo {author} {\bibfnamefont {H.}~\bibnamefont {Katsura}},\
  }\bibfield  {title} {\bibinfo {title} {Onsager's scars in disordered spin
  chains},\ }\href {https://doi.org/10.1103/PhysRevLett.124.180604} {\bibfield
  {journal} {\bibinfo  {journal} {Phys. Rev. Lett.}\ }\textbf {\bibinfo
  {volume} {124}},\ \bibinfo {pages} {180604} (\bibinfo {year}
  {2020})}\BibitemShut {NoStop}%
\bibitem [{\citenamefont {Mark}\ and\ \citenamefont
  {Motrunich}(2020)}]{MM:2018}%
  \BibitemOpen
  \bibfield  {author} {\bibinfo {author} {\bibfnamefont {D.~K.}\ \bibnamefont
  {Mark}}\ and\ \bibinfo {author} {\bibfnamefont {O.~I.}\ \bibnamefont
  {Motrunich}},\ }\bibfield  {title} {\bibinfo {title}
  {$\ensuremath{\eta}$-pairing states as true scars in an extended {Hubbard}
  model},\ }\href {https://doi.org/10.1103/PhysRevB.102.075132} {\bibfield
  {journal} {\bibinfo  {journal} {Phys. Rev. B}\ }\textbf {\bibinfo {volume}
  {102}},\ \bibinfo {pages} {075132} (\bibinfo {year} {2020})}\BibitemShut
  {NoStop}%
\bibitem [{\citenamefont {Moudgalya}\ \emph
  {et~al.}(2020{\natexlab{a}})\citenamefont {Moudgalya}, \citenamefont
  {Bernevig},\ and\ \citenamefont {Regnault}}]{MB:2020}%
  \BibitemOpen
  \bibfield  {author} {\bibinfo {author} {\bibfnamefont {S.}~\bibnamefont
  {Moudgalya}}, \bibinfo {author} {\bibfnamefont {B.~A.}\ \bibnamefont
  {Bernevig}},\ and\ \bibinfo {author} {\bibfnamefont {N.}~\bibnamefont
  {Regnault}},\ }\bibfield  {title} {\bibinfo {title} {Quantum many-body scars
  in a {Landau} level on a thin torus},\ }\href
  {https://doi.org/10.1103/PhysRevB.102.195150} {\bibfield  {journal} {\bibinfo
   {journal} {Phys. Rev. B}\ }\textbf {\bibinfo {volume} {102}},\ \bibinfo
  {pages} {195150} (\bibinfo {year} {2020}{\natexlab{a}})}\BibitemShut
  {NoStop}%
\bibitem [{\citenamefont {McClarty}\ \emph {et~al.}(2020)\citenamefont
  {McClarty}, \citenamefont {Haque}, \citenamefont {Sen},\ and\ \citenamefont
  {Richter}}]{MHSR:2020}%
  \BibitemOpen
  \bibfield  {author} {\bibinfo {author} {\bibfnamefont {P.~A.}\ \bibnamefont
  {McClarty}}, \bibinfo {author} {\bibfnamefont {M.}~\bibnamefont {Haque}},
  \bibinfo {author} {\bibfnamefont {A.}~\bibnamefont {Sen}},\ and\ \bibinfo
  {author} {\bibfnamefont {J.}~\bibnamefont {Richter}},\ }\bibfield  {title}
  {\bibinfo {title} {Disorder-free localization and many-body quantum scars
  from magnetic frustration},\ }\href
  {https://doi.org/10.1103/PhysRevB.102.224303} {\bibfield  {journal} {\bibinfo
   {journal} {Phys. Rev. B}\ }\textbf {\bibinfo {volume} {102}},\ \bibinfo
  {pages} {224303} (\bibinfo {year} {2020})}\BibitemShut {NoStop}%
\bibitem [{\citenamefont {Moudgalya}\ \emph
  {et~al.}(2020{\natexlab{b}})\citenamefont {Moudgalya}, \citenamefont
  {Regnault},\ and\ \citenamefont {Bernevig}}]{MRB:2020}%
  \BibitemOpen
  \bibfield  {author} {\bibinfo {author} {\bibfnamefont {S.}~\bibnamefont
  {Moudgalya}}, \bibinfo {author} {\bibfnamefont {N.}~\bibnamefont
  {Regnault}},\ and\ \bibinfo {author} {\bibfnamefont {B.~A.}\ \bibnamefont
  {Bernevig}},\ }\bibfield  {title} {\bibinfo {title}
  {$\ensuremath{\eta}$-pairing in {Hubbard} models: From spectrum generating
  algebras to quantum many-body scars},\ }\href
  {https://doi.org/10.1103/PhysRevB.102.085140} {\bibfield  {journal} {\bibinfo
   {journal} {Phys. Rev. B}\ }\textbf {\bibinfo {volume} {102}},\ \bibinfo
  {pages} {085140} (\bibinfo {year} {2020}{\natexlab{b}})}\BibitemShut
  {NoStop}%
\bibitem [{\citenamefont {van Voorden}\ \emph {et~al.}(2020)\citenamefont {van
  Voorden}, \citenamefont {Min\'a\ifmmode~\check{r}\else \v{r}\fi{}},\ and\
  \citenamefont {Schoutens}}]{VMS:2020}%
  \BibitemOpen
  \bibfield  {author} {\bibinfo {author} {\bibfnamefont {B.}~\bibnamefont {van
  Voorden}}, \bibinfo {author} {\bibfnamefont {J.~c.~v.}\ \bibnamefont
  {Min\'a\ifmmode~\check{r}\else \v{r}\fi{}}},\ and\ \bibinfo {author}
  {\bibfnamefont {K.}~\bibnamefont {Schoutens}},\ }\bibfield  {title} {\bibinfo
  {title} {Quantum many-body scars in transverse field {Ising} ladders and
  beyond},\ }\href {https://doi.org/10.1103/PhysRevB.101.220305} {\bibfield
  {journal} {\bibinfo  {journal} {Phys. Rev. B}\ }\textbf {\bibinfo {volume}
  {101}},\ \bibinfo {pages} {220305} (\bibinfo {year} {2020})}\BibitemShut
  {NoStop}%
\bibitem [{\citenamefont {Lin}\ \emph {et~al.}(2020)\citenamefont {Lin},
  \citenamefont {Calvera},\ and\ \citenamefont {Hsieh}}]{LCH:2020}%
  \BibitemOpen
  \bibfield  {author} {\bibinfo {author} {\bibfnamefont {C.-J.}\ \bibnamefont
  {Lin}}, \bibinfo {author} {\bibfnamefont {V.}~\bibnamefont {Calvera}},\ and\
  \bibinfo {author} {\bibfnamefont {T.~H.}\ \bibnamefont {Hsieh}},\ }\bibfield
  {title} {\bibinfo {title} {Quantum many-body scar states in two-dimensional
  {Rydberg} atom arrays},\ }\href {https://doi.org/10.1103/PhysRevB.101.220304}
  {\bibfield  {journal} {\bibinfo  {journal} {Phys. Rev. B}\ }\textbf {\bibinfo
  {volume} {101}},\ \bibinfo {pages} {220304} (\bibinfo {year}
  {2020})}\BibitemShut {NoStop}%
\bibitem [{\citenamefont {Hart}\ \emph {et~al.}(2020)\citenamefont {Hart},
  \citenamefont {De~Tomasi},\ and\ \citenamefont {Castelnovo}}]{HDGC:2020}%
  \BibitemOpen
  \bibfield  {author} {\bibinfo {author} {\bibfnamefont {O.}~\bibnamefont
  {Hart}}, \bibinfo {author} {\bibfnamefont {G.}~\bibnamefont {De~Tomasi}},\
  and\ \bibinfo {author} {\bibfnamefont {C.}~\bibnamefont {Castelnovo}},\
  }\bibfield  {title} {\bibinfo {title} {From compact localized states to
  many-body scars in the random quantum comb},\ }\href
  {https://doi.org/10.1103/PhysRevResearch.2.043267} {\bibfield  {journal}
  {\bibinfo  {journal} {Phys. Rev. Research}\ }\textbf {\bibinfo {volume}
  {2}},\ \bibinfo {pages} {043267} (\bibinfo {year} {2020})}\BibitemShut
  {NoStop}%
\bibitem [{\citenamefont {Lee}\ \emph {et~al.}(2020)\citenamefont {Lee},
  \citenamefont {Melendrez}, \citenamefont {Pal},\ and\ \citenamefont
  {Changlani}}]{LMPC:2020}%
  \BibitemOpen
  \bibfield  {author} {\bibinfo {author} {\bibfnamefont {K.}~\bibnamefont
  {Lee}}, \bibinfo {author} {\bibfnamefont {R.}~\bibnamefont {Melendrez}},
  \bibinfo {author} {\bibfnamefont {A.}~\bibnamefont {Pal}},\ and\ \bibinfo
  {author} {\bibfnamefont {H.~J.}\ \bibnamefont {Changlani}},\ }\bibfield
  {title} {\bibinfo {title} {Exact three-colored quantum scars from geometric
  frustration},\ }\href {https://doi.org/10.1103/PhysRevB.101.241111}
  {\bibfield  {journal} {\bibinfo  {journal} {Phys. Rev. B}\ }\textbf {\bibinfo
  {volume} {101}},\ \bibinfo {pages} {241111} (\bibinfo {year}
  {2020})}\BibitemShut {NoStop}%
\bibitem [{\citenamefont {Zhao}\ \emph {et~al.}(2020)\citenamefont {Zhao},
  \citenamefont {Vovrosh}, \citenamefont {Mintert},\ and\ \citenamefont
  {Knolle}}]{Zhao2020}%
  \BibitemOpen
  \bibfield  {author} {\bibinfo {author} {\bibfnamefont {H.}~\bibnamefont
  {Zhao}}, \bibinfo {author} {\bibfnamefont {J.}~\bibnamefont {Vovrosh}},
  \bibinfo {author} {\bibfnamefont {F.}~\bibnamefont {Mintert}},\ and\ \bibinfo
  {author} {\bibfnamefont {J.}~\bibnamefont {Knolle}},\ }\bibfield  {title}
  {\bibinfo {title} {Quantum many-body scars in optical lattices},\ }\href
  {https://doi.org/10.1103/PhysRevLett.124.160604} {\bibfield  {journal}
  {\bibinfo  {journal} {Phys. Rev. Lett.}\ }\textbf {\bibinfo {volume} {124}},\
  \bibinfo {pages} {160604} (\bibinfo {year} {2020})}\BibitemShut {NoStop}%
\bibitem [{\citenamefont {Kuno}\ \emph {et~al.}(2020)\citenamefont {Kuno},
  \citenamefont {Mizoguchi},\ and\ \citenamefont {Hatsugai}}]{KMH:2020}%
  \BibitemOpen
  \bibfield  {author} {\bibinfo {author} {\bibfnamefont {Y.}~\bibnamefont
  {Kuno}}, \bibinfo {author} {\bibfnamefont {T.}~\bibnamefont {Mizoguchi}},\
  and\ \bibinfo {author} {\bibfnamefont {Y.}~\bibnamefont {Hatsugai}},\
  }\bibfield  {title} {\bibinfo {title} {Flat band quantum scar},\ }\href
  {https://doi.org/10.1103/PhysRevB.102.241115} {\bibfield  {journal} {\bibinfo
   {journal} {Phys. Rev. B}\ }\textbf {\bibinfo {volume} {102}},\ \bibinfo
  {pages} {241115} (\bibinfo {year} {2020})}\BibitemShut {NoStop}%
\bibitem [{\citenamefont {Mukherjee}\ \emph {et~al.}(2020)\citenamefont
  {Mukherjee}, \citenamefont {Nandy}, \citenamefont {Sen}, \citenamefont
  {Sen},\ and\ \citenamefont {Sengupta}}]{MNSS:2020}%
  \BibitemOpen
  \bibfield  {author} {\bibinfo {author} {\bibfnamefont {B.}~\bibnamefont
  {Mukherjee}}, \bibinfo {author} {\bibfnamefont {S.}~\bibnamefont {Nandy}},
  \bibinfo {author} {\bibfnamefont {A.}~\bibnamefont {Sen}}, \bibinfo {author}
  {\bibfnamefont {D.}~\bibnamefont {Sen}},\ and\ \bibinfo {author}
  {\bibfnamefont {K.}~\bibnamefont {Sengupta}},\ }\bibfield  {title} {\bibinfo
  {title} {Collapse and revival of quantum many-body scars via {Floquet}
  engineering},\ }\href {https://doi.org/10.1103/PhysRevB.101.245107}
  {\bibfield  {journal} {\bibinfo  {journal} {Phys. Rev. B}\ }\textbf {\bibinfo
  {volume} {101}},\ \bibinfo {pages} {245107} (\bibinfo {year}
  {2020})}\BibitemShut {NoStop}%
\bibitem [{\citenamefont {O'Dea}\ \emph {et~al.}(2020)\citenamefont {O'Dea},
  \citenamefont {Burnell}, \citenamefont {Chandran},\ and\ \citenamefont
  {Khemani}}]{ODea2020}%
  \BibitemOpen
  \bibfield  {author} {\bibinfo {author} {\bibfnamefont {N.}~\bibnamefont
  {O'Dea}}, \bibinfo {author} {\bibfnamefont {F.}~\bibnamefont {Burnell}},
  \bibinfo {author} {\bibfnamefont {A.}~\bibnamefont {Chandran}},\ and\
  \bibinfo {author} {\bibfnamefont {V.}~\bibnamefont {Khemani}},\ }\bibfield
  {title} {\bibinfo {title} {From tunnels to towers: Quantum scars from lie
  algebras and $q$-deformed lie algebras},\ }\href
  {https://doi.org/10.1103/PhysRevResearch.2.043305} {\bibfield  {journal}
  {\bibinfo  {journal} {Phys. Rev. Research}\ }\textbf {\bibinfo {volume}
  {2}},\ \bibinfo {pages} {043305} (\bibinfo {year} {2020})}\BibitemShut
  {NoStop}%
\bibitem [{\citenamefont {Surace}\ \emph {et~al.}(2021)\citenamefont {Surace},
  \citenamefont {Votto}, \citenamefont {Lazo}, \citenamefont {Silva},
  \citenamefont {Dalmonte},\ and\ \citenamefont {Giudici}}]{SVLSDG:2021}%
  \BibitemOpen
  \bibfield  {author} {\bibinfo {author} {\bibfnamefont {F.~M.}\ \bibnamefont
  {Surace}}, \bibinfo {author} {\bibfnamefont {M.}~\bibnamefont {Votto}},
  \bibinfo {author} {\bibfnamefont {E.~G.}\ \bibnamefont {Lazo}}, \bibinfo
  {author} {\bibfnamefont {A.}~\bibnamefont {Silva}}, \bibinfo {author}
  {\bibfnamefont {M.}~\bibnamefont {Dalmonte}},\ and\ \bibinfo {author}
  {\bibfnamefont {G.}~\bibnamefont {Giudici}},\ }\bibfield  {title} {\bibinfo
  {title} {Exact many-body scars and their stability in constrained quantum
  chains},\ }\href {https://doi.org/10.1103/PhysRevB.103.104302} {\bibfield
  {journal} {\bibinfo  {journal} {Phys. Rev. B}\ }\textbf {\bibinfo {volume}
  {103}},\ \bibinfo {pages} {104302} (\bibinfo {year} {2021})}\BibitemShut
  {NoStop}%
\bibitem [{\citenamefont {Wildeboer}\ \emph {et~al.}(2021)\citenamefont
  {Wildeboer}, \citenamefont {Seidel}, \citenamefont {Srivatsa}, \citenamefont
  {Nielsen},\ and\ \citenamefont {Erten}}]{WSSN:2021}%
  \BibitemOpen
  \bibfield  {author} {\bibinfo {author} {\bibfnamefont {J.}~\bibnamefont
  {Wildeboer}}, \bibinfo {author} {\bibfnamefont {A.}~\bibnamefont {Seidel}},
  \bibinfo {author} {\bibfnamefont {N.~S.}\ \bibnamefont {Srivatsa}}, \bibinfo
  {author} {\bibfnamefont {A.~E.~B.}\ \bibnamefont {Nielsen}},\ and\ \bibinfo
  {author} {\bibfnamefont {O.}~\bibnamefont {Erten}},\ }\bibfield  {title}
  {\bibinfo {title} {Topological quantum many-body scars in quantum dimer
  models on the kagome lattice},\ }\href
  {https://doi.org/10.1103/PhysRevB.104.L121103} {\bibfield  {journal}
  {\bibinfo  {journal} {Phys. Rev. B}\ }\textbf {\bibinfo {volume} {104}},\
  \bibinfo {pages} {L121103} (\bibinfo {year} {2021})}\BibitemShut {NoStop}%
\bibitem [{\citenamefont {Desaules}\ \emph {et~al.}(2021)\citenamefont
  {Desaules}, \citenamefont {Hudomal}, \citenamefont {Turner},\ and\
  \citenamefont {Papi\ifmmode~\acute{c}\else \'{c}\fi{}}}]{DHTP:2021}%
  \BibitemOpen
  \bibfield  {author} {\bibinfo {author} {\bibfnamefont {J.-Y.}\ \bibnamefont
  {Desaules}}, \bibinfo {author} {\bibfnamefont {A.}~\bibnamefont {Hudomal}},
  \bibinfo {author} {\bibfnamefont {C.~J.}\ \bibnamefont {Turner}},\ and\
  \bibinfo {author} {\bibfnamefont {Z.}~\bibnamefont
  {Papi\ifmmode~\acute{c}\else \'{c}\fi{}}},\ }\bibfield  {title} {\bibinfo
  {title} {Proposal for realizing quantum scars in the tilted {1D
  Fermi-Hubbard} model},\ }\href
  {https://doi.org/10.1103/PhysRevLett.126.210601} {\bibfield  {journal}
  {\bibinfo  {journal} {Phys. Rev. Lett.}\ }\textbf {\bibinfo {volume} {126}},\
  \bibinfo {pages} {210601} (\bibinfo {year} {2021})}\BibitemShut {NoStop}%
\bibitem [{\citenamefont {Ren}\ \emph {et~al.}(2021)\citenamefont {Ren},
  \citenamefont {Liang},\ and\ \citenamefont {Fang}}]{RLF:2021}%
  \BibitemOpen
  \bibfield  {author} {\bibinfo {author} {\bibfnamefont {J.}~\bibnamefont
  {Ren}}, \bibinfo {author} {\bibfnamefont {C.}~\bibnamefont {Liang}},\ and\
  \bibinfo {author} {\bibfnamefont {C.}~\bibnamefont {Fang}},\ }\bibfield
  {title} {\bibinfo {title} {Quasisymmetry groups and many-body scar
  dynamics},\ }\href {https://doi.org/10.1103/PhysRevLett.126.120604}
  {\bibfield  {journal} {\bibinfo  {journal} {Phys. Rev. Lett.}\ }\textbf
  {\bibinfo {volume} {126}},\ \bibinfo {pages} {120604} (\bibinfo {year}
  {2021})}\BibitemShut {NoStop}%
\bibitem [{\citenamefont {Fendley}\ \emph {et~al.}(2004)\citenamefont
  {Fendley}, \citenamefont {Sengupta},\ and\ \citenamefont
  {Sachdev}}]{FendleySachdev}%
  \BibitemOpen
  \bibfield  {author} {\bibinfo {author} {\bibfnamefont {P.}~\bibnamefont
  {Fendley}}, \bibinfo {author} {\bibfnamefont {K.}~\bibnamefont {Sengupta}},\
  and\ \bibinfo {author} {\bibfnamefont {S.}~\bibnamefont {Sachdev}},\
  }\bibfield  {title} {\bibinfo {title} {Competing density-wave orders in a
  one-dimensional hard-boson model},\ }\href
  {https://doi.org/10.1103/PhysRevB.69.075106} {\bibfield  {journal} {\bibinfo
  {journal} {Phys. Rev. B}\ }\textbf {\bibinfo {volume} {69}},\ \bibinfo
  {pages} {075106} (\bibinfo {year} {2004})}\BibitemShut {NoStop}%
\bibitem [{\citenamefont {Lesanovsky}\ and\ \citenamefont
  {Katsura}(2012)}]{Lesanovsky2012}%
  \BibitemOpen
  \bibfield  {author} {\bibinfo {author} {\bibfnamefont {I.}~\bibnamefont
  {Lesanovsky}}\ and\ \bibinfo {author} {\bibfnamefont {H.}~\bibnamefont
  {Katsura}},\ }\bibfield  {title} {\bibinfo {title} {Interacting {Fibonacci}
  anyons in a {Rydberg} gas},\ }\href
  {https://doi.org/10.1103/PhysRevA.86.041601} {\bibfield  {journal} {\bibinfo
  {journal} {Phys. Rev. A}\ }\textbf {\bibinfo {volume} {86}},\ \bibinfo
  {pages} {041601} (\bibinfo {year} {2012})}\BibitemShut {NoStop}%
\bibitem [{\citenamefont {Bernien}\ \emph {et~al.}(2017)\citenamefont
  {Bernien}, \citenamefont {Schwartz}, \citenamefont {Keesling}, \citenamefont
  {Levine}, \citenamefont {Omran}, \citenamefont {Pichler}, \citenamefont
  {Choi}, \citenamefont {Zibrov}, \citenamefont {Endres}, \citenamefont
  {Greiner}, \citenamefont {Vuleti{\'{c}}},\ and\ \citenamefont
  {Lukin}}]{BSKL:2017}%
  \BibitemOpen
  \bibfield  {author} {\bibinfo {author} {\bibfnamefont {H.}~\bibnamefont
  {Bernien}}, \bibinfo {author} {\bibfnamefont {S.}~\bibnamefont {Schwartz}},
  \bibinfo {author} {\bibfnamefont {A.}~\bibnamefont {Keesling}}, \bibinfo
  {author} {\bibfnamefont {H.}~\bibnamefont {Levine}}, \bibinfo {author}
  {\bibfnamefont {A.}~\bibnamefont {Omran}}, \bibinfo {author} {\bibfnamefont
  {H.}~\bibnamefont {Pichler}}, \bibinfo {author} {\bibfnamefont
  {S.}~\bibnamefont {Choi}}, \bibinfo {author} {\bibfnamefont {A.~S.}\
  \bibnamefont {Zibrov}}, \bibinfo {author} {\bibfnamefont {M.}~\bibnamefont
  {Endres}}, \bibinfo {author} {\bibfnamefont {M.}~\bibnamefont {Greiner}},
  \bibinfo {author} {\bibfnamefont {V.}~\bibnamefont {Vuleti{\'{c}}}},\ and\
  \bibinfo {author} {\bibfnamefont {M.~D.}\ \bibnamefont {Lukin}},\ }\bibfield
  {title} {\bibinfo {title} {Probing many-body dynamics on a 51-atom quantum
  simulator},\ }\href {https://doi.org/10.1038/nature24622} {\bibfield
  {journal} {\bibinfo  {journal} {Nature}\ }\textbf {\bibinfo {volume} {551}},\
  \bibinfo {pages} {579} (\bibinfo {year} {2017})}\BibitemShut {NoStop}%
\bibitem [{\citenamefont {Su}\ \emph {et~al.}(2022)\citenamefont {Su},
  \citenamefont {Sun}, \citenamefont {Hudomal}, \citenamefont {Desaules},
  \citenamefont {Zhou}, \citenamefont {Yang}, \citenamefont {Halimeh},
  \citenamefont {Yuan}, \citenamefont {Papić},\ and\ \citenamefont
  {Pan}}]{Su2022}%
  \BibitemOpen
  \bibfield  {author} {\bibinfo {author} {\bibfnamefont {G.-X.}\ \bibnamefont
  {Su}}, \bibinfo {author} {\bibfnamefont {H.}~\bibnamefont {Sun}}, \bibinfo
  {author} {\bibfnamefont {A.}~\bibnamefont {Hudomal}}, \bibinfo {author}
  {\bibfnamefont {J.-Y.}\ \bibnamefont {Desaules}}, \bibinfo {author}
  {\bibfnamefont {Z.-Y.}\ \bibnamefont {Zhou}}, \bibinfo {author}
  {\bibfnamefont {B.}~\bibnamefont {Yang}}, \bibinfo {author} {\bibfnamefont
  {J.~C.}\ \bibnamefont {Halimeh}}, \bibinfo {author} {\bibfnamefont {Z.-S.}\
  \bibnamefont {Yuan}}, \bibinfo {author} {\bibfnamefont {Z.}~\bibnamefont
  {Papić}},\ and\ \bibinfo {author} {\bibfnamefont {J.-W.}\ \bibnamefont
  {Pan}},\ }\href@noop {} {\bibinfo {title} {Observation of unconventional
  many-body scarring in a quantum simulator}} (\bibinfo {year} {2022}),\
  \Eprint {https://arxiv.org/abs/2201.00821} {arXiv:2201.00821
  [cond-mat.quant-gas]} \BibitemShut {NoStop}%
\bibitem [{\citenamefont {Jepsen}\ \emph {et~al.}(2021)\citenamefont {Jepsen},
  \citenamefont {Lee}, \citenamefont {Lin}, \citenamefont {Dimitrova},
  \citenamefont {Margalit}, \citenamefont {Ho},\ and\ \citenamefont
  {Ketterle}}]{Jepsen2021}%
  \BibitemOpen
  \bibfield  {author} {\bibinfo {author} {\bibfnamefont {P.~N.}\ \bibnamefont
  {Jepsen}}, \bibinfo {author} {\bibfnamefont {Y.~K.}\ \bibnamefont {Lee}},
  \bibinfo {author} {\bibfnamefont {H.}~\bibnamefont {Lin}}, \bibinfo {author}
  {\bibfnamefont {I.}~\bibnamefont {Dimitrova}}, \bibinfo {author}
  {\bibfnamefont {Y.}~\bibnamefont {Margalit}}, \bibinfo {author}
  {\bibfnamefont {W.~W.}\ \bibnamefont {Ho}},\ and\ \bibinfo {author}
  {\bibfnamefont {W.}~\bibnamefont {Ketterle}},\ }\href@noop {} {\bibinfo
  {title} {Catching {Bethe} phantoms and quantum many-body scars: Long-lived
  spin-helix states in {Heisenberg} magnets}} (\bibinfo {year} {2021}),\
  \Eprint {https://arxiv.org/abs/2110.12043} {arXiv:2110.12043
  [cond-mat.quant-gas]} \BibitemShut {NoStop}%
\bibitem [{\citenamefont {Su}\ \emph {et~al.}(1979)\citenamefont {Su},
  \citenamefont {Schrieffer},\ and\ \citenamefont {Heeger}}]{SSHModel}%
  \BibitemOpen
  \bibfield  {author} {\bibinfo {author} {\bibfnamefont {W.~P.}\ \bibnamefont
  {Su}}, \bibinfo {author} {\bibfnamefont {J.~R.}\ \bibnamefont {Schrieffer}},\
  and\ \bibinfo {author} {\bibfnamefont {A.~J.}\ \bibnamefont {Heeger}},\
  }\bibfield  {title} {\bibinfo {title} {Solitons in polyacetylene},\ }\href
  {https://doi.org/10.1103/PhysRevLett.42.1698} {\bibfield  {journal} {\bibinfo
   {journal} {Phys. Rev. Lett.}\ }\textbf {\bibinfo {volume} {42}},\ \bibinfo
  {pages} {1698} (\bibinfo {year} {1979})}\BibitemShut {NoStop}%
\bibitem [{\citenamefont {Wu}\ \emph {et~al.}(2021)\citenamefont {Wu},
  \citenamefont {Bao}, \citenamefont {Cao}, \citenamefont {Chen}, \citenamefont
  {Chen}, \citenamefont {Chen}, \citenamefont {Chung}, \citenamefont {Deng},
  \citenamefont {Du}, \citenamefont {Fan} \emph {et~al.}}]{WBCC:2021}%
  \BibitemOpen
  \bibfield  {author} {\bibinfo {author} {\bibfnamefont {Y.}~\bibnamefont
  {Wu}}, \bibinfo {author} {\bibfnamefont {W.-S.}\ \bibnamefont {Bao}},
  \bibinfo {author} {\bibfnamefont {S.}~\bibnamefont {Cao}}, \bibinfo {author}
  {\bibfnamefont {F.}~\bibnamefont {Chen}}, \bibinfo {author} {\bibfnamefont
  {M.-C.}\ \bibnamefont {Chen}}, \bibinfo {author} {\bibfnamefont
  {X.}~\bibnamefont {Chen}}, \bibinfo {author} {\bibfnamefont {T.-H.}\
  \bibnamefont {Chung}}, \bibinfo {author} {\bibfnamefont {H.}~\bibnamefont
  {Deng}}, \bibinfo {author} {\bibfnamefont {Y.}~\bibnamefont {Du}}, \bibinfo
  {author} {\bibfnamefont {D.}~\bibnamefont {Fan}}, \emph {et~al.},\ }\bibfield
   {title} {\bibinfo {title} {Strong quantum computational advantage using a
  superconducting quantum processor},\ }\href
  {https://doi.org/10.1103/PhysRevLett.127.180501} {\bibfield  {journal}
  {\bibinfo  {journal} {arXiv preprint arXiv:2106.14734}\ } (\bibinfo {year}
  {2021})}\BibitemShut {NoStop}%
\bibitem [{\citenamefont {Krantz}\ \emph {et~al.}(2019)\citenamefont {Krantz},
  \citenamefont {Kjaergaard}, \citenamefont {Yan}, \citenamefont {Orlando},
  \citenamefont {Gustavsson},\ and\ \citenamefont {Oliver}}]{KKYO:2019}%
  \BibitemOpen
  \bibfield  {author} {\bibinfo {author} {\bibfnamefont {P.}~\bibnamefont
  {Krantz}}, \bibinfo {author} {\bibfnamefont {M.}~\bibnamefont {Kjaergaard}},
  \bibinfo {author} {\bibfnamefont {F.}~\bibnamefont {Yan}}, \bibinfo {author}
  {\bibfnamefont {T.~P.}\ \bibnamefont {Orlando}}, \bibinfo {author}
  {\bibfnamefont {S.}~\bibnamefont {Gustavsson}},\ and\ \bibinfo {author}
  {\bibfnamefont {W.~D.}\ \bibnamefont {Oliver}},\ }\bibfield  {title}
  {\bibinfo {title} {A quantum engineer's guide to superconducting qubits},\
  }\href {https://doi.org/10.1063/1.5089550} {\bibfield  {journal} {\bibinfo
  {journal} {Applied Physics Reviews}\ }\textbf {\bibinfo {volume} {6}},\
  \bibinfo {pages} {021318} (\bibinfo {year} {2019})}\BibitemShut {NoStop}%
\bibitem [{\citenamefont {Blais}\ \emph {et~al.}(2021)\citenamefont {Blais},
  \citenamefont {Grimsmo}, \citenamefont {Girvin},\ and\ \citenamefont
  {Wallraff}}]{BGGW:2021}%
  \BibitemOpen
  \bibfield  {author} {\bibinfo {author} {\bibfnamefont {A.}~\bibnamefont
  {Blais}}, \bibinfo {author} {\bibfnamefont {A.~L.}\ \bibnamefont {Grimsmo}},
  \bibinfo {author} {\bibfnamefont {S.~M.}\ \bibnamefont {Girvin}},\ and\
  \bibinfo {author} {\bibfnamefont {A.}~\bibnamefont {Wallraff}},\ }\bibfield
  {title} {\bibinfo {title} {Circuit quantum electrodynamics},\ }\href
  {https://doi.org/10.1103/RevModPhys.93.025005} {\bibfield  {journal}
  {\bibinfo  {journal} {Rev. Mod. Phys.}\ }\textbf {\bibinfo {volume} {93}},\
  \bibinfo {pages} {025005} (\bibinfo {year} {2021})}\BibitemShut {NoStop}%
\bibitem [{\citenamefont {Jafari}\ and\ \citenamefont
  {Johannesson}(2017)}]{Jafari2017}%
  \BibitemOpen
  \bibfield  {author} {\bibinfo {author} {\bibfnamefont {R.}~\bibnamefont
  {Jafari}}\ and\ \bibinfo {author} {\bibfnamefont {H.}~\bibnamefont
  {Johannesson}},\ }\bibfield  {title} {\bibinfo {title} {Loschmidt echo
  revivals: Critical and noncritical},\ }\href
  {https://doi.org/10.1103/PhysRevLett.118.015701} {\bibfield  {journal}
  {\bibinfo  {journal} {Phys. Rev. Lett.}\ }\textbf {\bibinfo {volume} {118}},\
  \bibinfo {pages} {015701} (\bibinfo {year} {2017})}\BibitemShut {NoStop}%
\bibitem [{\citenamefont {Najafi}\ \emph {et~al.}(2019)\citenamefont {Najafi},
  \citenamefont {Rajabpour},\ and\ \citenamefont {Viti}}]{Najafi2019}%
  \BibitemOpen
  \bibfield  {author} {\bibinfo {author} {\bibfnamefont {K.}~\bibnamefont
  {Najafi}}, \bibinfo {author} {\bibfnamefont {M.~A.}\ \bibnamefont
  {Rajabpour}},\ and\ \bibinfo {author} {\bibfnamefont {J.}~\bibnamefont
  {Viti}},\ }\bibfield  {title} {\bibinfo {title} {Return amplitude after a
  quantum quench in the {XY} chain},\ }\href
  {https://doi.org/10.1088/1742-5468/ab3413} {\bibfield  {journal} {\bibinfo
  {journal} {Journal of Statistical Mechanics: Theory and Experiment}\ }\textbf
  {\bibinfo {volume} {2019}},\ \bibinfo {pages} {083102} (\bibinfo {year}
  {2019})}\BibitemShut {NoStop}%
\bibitem [{\citenamefont {Maimaiti}\ \emph {et~al.}(2017)\citenamefont
  {Maimaiti}, \citenamefont {Andreanov}, \citenamefont {Park}, \citenamefont
  {Gendelman},\ and\ \citenamefont {Flach}}]{Maimaiti2017}%
  \BibitemOpen
  \bibfield  {author} {\bibinfo {author} {\bibfnamefont {W.}~\bibnamefont
  {Maimaiti}}, \bibinfo {author} {\bibfnamefont {A.}~\bibnamefont {Andreanov}},
  \bibinfo {author} {\bibfnamefont {H.~C.}\ \bibnamefont {Park}}, \bibinfo
  {author} {\bibfnamefont {O.}~\bibnamefont {Gendelman}},\ and\ \bibinfo
  {author} {\bibfnamefont {S.}~\bibnamefont {Flach}},\ }\bibfield  {title}
  {\bibinfo {title} {Compact localized states and flat-band generators in one
  dimension},\ }\href {https://doi.org/10.1103/PhysRevB.95.115135} {\bibfield
  {journal} {\bibinfo  {journal} {Phys. Rev. B}\ }\textbf {\bibinfo {volume}
  {95}},\ \bibinfo {pages} {115135} (\bibinfo {year} {2017})}\BibitemShut
  {NoStop}%
\bibitem [{\citenamefont {Xia}\ \emph {et~al.}(2021)\citenamefont {Xia},
  \citenamefont {Zou}, \citenamefont {Qiu},\ and\ \citenamefont
  {Li}}]{XZQL:2021}%
  \BibitemOpen
  \bibfield  {author} {\bibinfo {author} {\bibfnamefont {W.}~\bibnamefont
  {Xia}}, \bibinfo {author} {\bibfnamefont {J.}~\bibnamefont {Zou}}, \bibinfo
  {author} {\bibfnamefont {X.}~\bibnamefont {Qiu}},\ and\ \bibinfo {author}
  {\bibfnamefont {X.}~\bibnamefont {Li}},\ }\href@noop {} {\bibinfo {title}
  {The reservoir learning power across quantum many-boby localization
  transition}} (\bibinfo {year} {2021}),\ \Eprint
  {https://arxiv.org/abs/2104.02727} {arXiv:2104.02727 [quant-ph]} \BibitemShut
  {NoStop}%
\bibitem [{\citenamefont {Neill}\ \emph {et~al.}(2018)\citenamefont {Neill},
  \citenamefont {Roushan}, \citenamefont {Kechedzhi}, \citenamefont {Boixo},
  \citenamefont {Isakov}, \citenamefont {Smelyanskiy}, \citenamefont {Megrant},
  \citenamefont {Chiaro}, \citenamefont {Dunsworth}, \citenamefont {Arya} \emph
  {et~al.}}]{NRKB:2018}%
  \BibitemOpen
  \bibfield  {author} {\bibinfo {author} {\bibfnamefont {C.}~\bibnamefont
  {Neill}}, \bibinfo {author} {\bibfnamefont {P.}~\bibnamefont {Roushan}},
  \bibinfo {author} {\bibfnamefont {K.}~\bibnamefont {Kechedzhi}}, \bibinfo
  {author} {\bibfnamefont {S.}~\bibnamefont {Boixo}}, \bibinfo {author}
  {\bibfnamefont {S.~V.}\ \bibnamefont {Isakov}}, \bibinfo {author}
  {\bibfnamefont {V.}~\bibnamefont {Smelyanskiy}}, \bibinfo {author}
  {\bibfnamefont {A.}~\bibnamefont {Megrant}}, \bibinfo {author} {\bibfnamefont
  {B.}~\bibnamefont {Chiaro}}, \bibinfo {author} {\bibfnamefont
  {A.}~\bibnamefont {Dunsworth}}, \bibinfo {author} {\bibfnamefont
  {K.}~\bibnamefont {Arya}}, \emph {et~al.},\ }\bibfield  {title} {\bibinfo
  {title} {A blueprint for demonstrating quantum supremacy with superconducting
  qubits},\ }\href {https://doi.org/10.1126/science.aao4309} {\bibfield
  {journal} {\bibinfo  {journal} {Science}\ }\textbf {\bibinfo {volume}
  {360}},\ \bibinfo {pages} {195} (\bibinfo {year} {2018})}\BibitemShut
  {NoStop}%
\end{thebibliography}

\end{document}